\documentclass[lettersize,journal]{IEEEtran}
\usepackage{amsmath,amsfonts}
\usepackage{algorithmic}
\usepackage{algorithm}
\usepackage{array}
\usepackage[caption=false,font=normalsize,labelfont=sf,textfont=sf]{subfig}
\usepackage{textcomp}
\usepackage{stfloats}
\usepackage{url}
\usepackage{verbatim}
\usepackage{graphicx}
\usepackage{cite}
\usepackage{multirow}
\usepackage{booktabs}
\usepackage{subfig}
\usepackage{marvosym}
\usepackage{xcolor}
\usepackage{tabularx}
\usepackage{hyperref} 
\usepackage{mdframed}
\definecolor{lightgray}{gray}{0.95}
\begin{document}

\title{FlowXpert: Context-Aware Flow Embedding for Enhanced Traffic Detection in IoT Network}


%
\author{\IEEEauthorblockN{Chao Zha,
Haolin Pan,
Bing Bai,
Jiangxing Wu,
and Ruyun Zhang\textsuperscript{\Letter}}

\thanks{This paper was produced by the IEEE Publication Technology Group. They are in Piscataway, NJ.}
\thanks{Manuscript received April 19, 2021; revised August 16, 2021 (Corresponding author: Ruyun Zhang).}
\thanks{Chao Zha is with the Institute of Computing Technology, Chinese Academy of Sciences, Beijing 100190, China; also with the Research Center for High Efficiency Computing Infrastructure, Zhejiang Lab, Hangzhou 311100, Zhejiang, China; and with the University of the Chinese Academy of Sciences, Beijing 100049, China. (email: \texttt{\href{mailto:zhachao21@mails.ucas.ac.cn}{zhachao21@mails.ucas.ac.cn}}).}

\thanks{Haolin Pan is with the School of Intelligent Science and Technology, Hangzhou Institute for Advanced Study, UCAS, Hangzhou 311100, Zhejiang, China; and with the University of the Chinese Academy of Sciences, Beijing 100049, China. (email: \texttt{\href{mailto:haolinpan21@mails.ucas.ac.cn}{panhaolin21@mails.ucas.ac.cn}}).}

\thanks{Bing Bai, and Ruyun Zhang are with the Research Center for High Efficiency Computing Infrastructure, Zhejiang Lab, Hangzhou 311100, Zhejiang, China (email: \texttt{\href{mailto:baibing@zhejianglab.org}{baibing@zhejianglab.org}}; \texttt{\href{mailto:zcor2021@gmail.com}{zcor2021@gmail.com}}).}

\thanks{Jiangxing Wu is with the National Digital Switching System Engineering and Technological R\&D Center, Zhengzhou 450003, China. (email: \texttt{\href{ndscwjx@126.com}{ndscwjx@126.com}}).}
}

\markboth{}%
{Shell \MakeLowercase{\textit{et al.}}: A Sample Article Using IEEEtran.cls for IEEE Journals}


\maketitle

\begin{abstract}
In the Internet of Things (IoT) environment, continuous interaction among a large number of devices generates complex and dynamic network traffic, which poses significant challenges to rule-based detection approaches. Machine learning (ML)-based traffic detection technology, capable of identifying anomalous patterns and potential threats within this traffic, serves as a critical component in ensuring network security. This study first identifies a significant issue with widely adopted feature extraction tools (e.g., CICMeterFlow): the extensive use of time- and length-related features leads to high sparsity, which adversely affects model convergence. Furthermore, existing traffic detection methods generally lack an embedding mechanism capable of efficiently and comprehensively capturing the semantic characteristics of network traffic. To address these challenges, we propose a novel feature extraction tool that eliminates traditional time and length features in favor of context-aware semantic features related to the source host, thus improving the generalizability of the model. In addition, we design an embedding training framework that integrates the unsupervised DBSCAN clustering algorithm with a contrastive learning strategy to effectively capture fine-grained semantic representations of traffic. Extensive empirical evaluations are conducted on the real-world Mawi data set to validate the proposed method in terms of detection accuracy, robustness, and generalization. Comparative experiments against several state-of-the-art (SOTA) models demonstrate the superior performance of our approach. Furthermore, we confirm its applicability and deployability in real-time scenarios.
\end{abstract}


\section{Introduction}
\IEEEPARstart{W}{ith} the rapid proliferation of the IoT, a huge number of interconnected devices are continuously generating and exchanging data over the network. This ubiquitous connectivity has significantly expanded the attack surface, making IoT networks a prime target for various cyber threats. Moreover, the resource-constrained nature of many IoT devices, along with the increasing use of encryption in IoT traffic, poses serious challenges to traditional network security mechanisms. Traditional traffic detection methods that rely on predefined rules or signature-based matching are often ineffective in identifying novel or sophisticated attacks in IoT environments \cite{costa2023rule, ilgun2002state, masdari2020survey}. As a result, there has been a growing research interest in using machine learning and deep learning techniques to develop more adaptive and intelligent traffic detection systems customized for IoT networks \cite{zha2025nids}.

Specifically, traditional machine learning algorithms, such as decision trees, random forests, and support vector machines, have been widely applied to network traffic detection tasks, particularly within the paradigm of supervised learning \cite{louk2023dual, ma2021novel, resende2018survey}. In parallel, unsupervised clustering algorithms have also been utilized to uncover latent patterns in traffic data \cite{fu2021realtime, fu2023detecting, wang2013internet, zhang2014robust}. Building on these foundations, extensive research has been conducted to further enhance detection performance. At the same time, deep learning models, including multilayer perceptrons, autoencoders, and conditional probabilistic models, have been introduced into this domain \cite{du2017deeplog, mirsky2018kitsune, yang2021conditional, zha2025nids,zha2025dm, zha2024skt, lin2025convolutions}. These models enable automatic extraction of salient features, which improves the efficiency and accuracy of traffic classification.

However, existing research efforts remain largely focused on achieving "idealized" performance metrics in benchmark or simulated datasets \cite{sommer2010outside}, with limited emphasis on conducting systematic analyzes of the intrusion detection task itself. Many studies tend to apply artificial intelligence techniques directly to network traffic detection without adequately incorporating the practical characteristics and contextual nuances of real-world network environments. As a result, truly effective and efficient detection capabilities have yet to be fully realized. Despite significant progress in the field, network traffic detection still faces numerous critical challenges that warrant further in-depth investigation and innovation.

\paragraph{\textbf{Challenge 1}} \textit{Limitations of Flow-Based Features in Model Convergence.} A network flow is typically defined as a set of packets that share the same five-tuple: source IP, destination IP, source port, destination port, and transport-layer protocol within a specific time window \cite{yang2021conditional}. Flow features are structured statistical attributes extracted from these packet sets and serve as the main input for model training in benchmark data sets such as NSL-KDD \cite{Tavallaee2009}, CICIDS-2017 \cite{CICIDS2017}, CICIDS-2018 \cite{CICIDS2018}, and UNSW-NB-15 \cite{moustafa2015unsw}. These features are generally categorized into temporal, packet length-related, and protocol-related types. Empirical evidence shows that their value distributions are often highly skewed and sparse, with most values concentrated near zero, as shown in Fig. 1. Most flows are short-lived, and most packets are small, while long-duration flows and large packets are rare. This asymmetry deviates from a standard Gaussian distribution and results in a highly sparse feature space after normalization.
\begin{figure}[ht]
\centering
\subfloat[NSL-KDD.]{\includegraphics[width=0.48\textwidth]{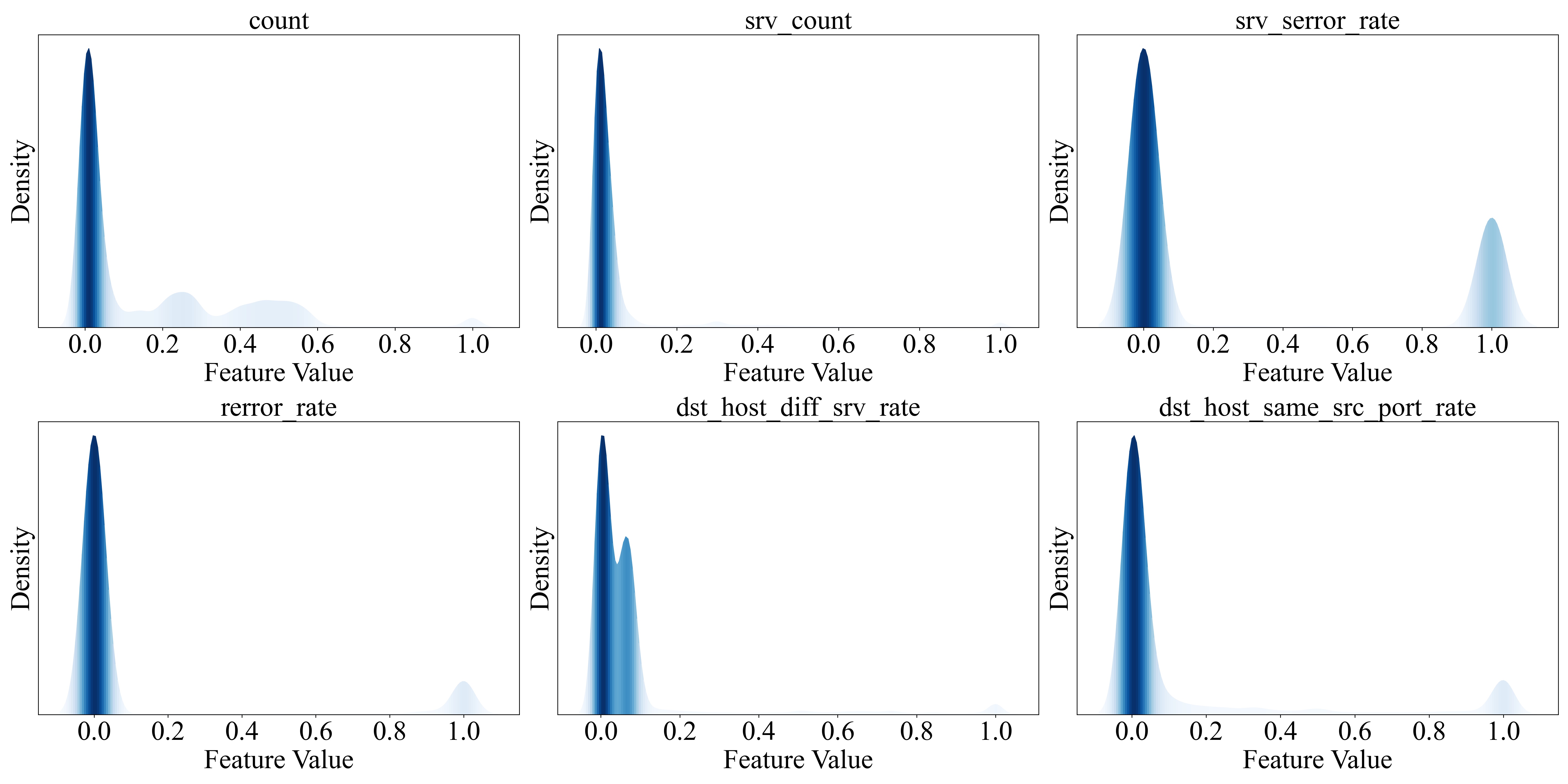}%
\label{fig_1-1a}}

\subfloat[UNSW-NB16.]{\includegraphics[width=0.48\textwidth]{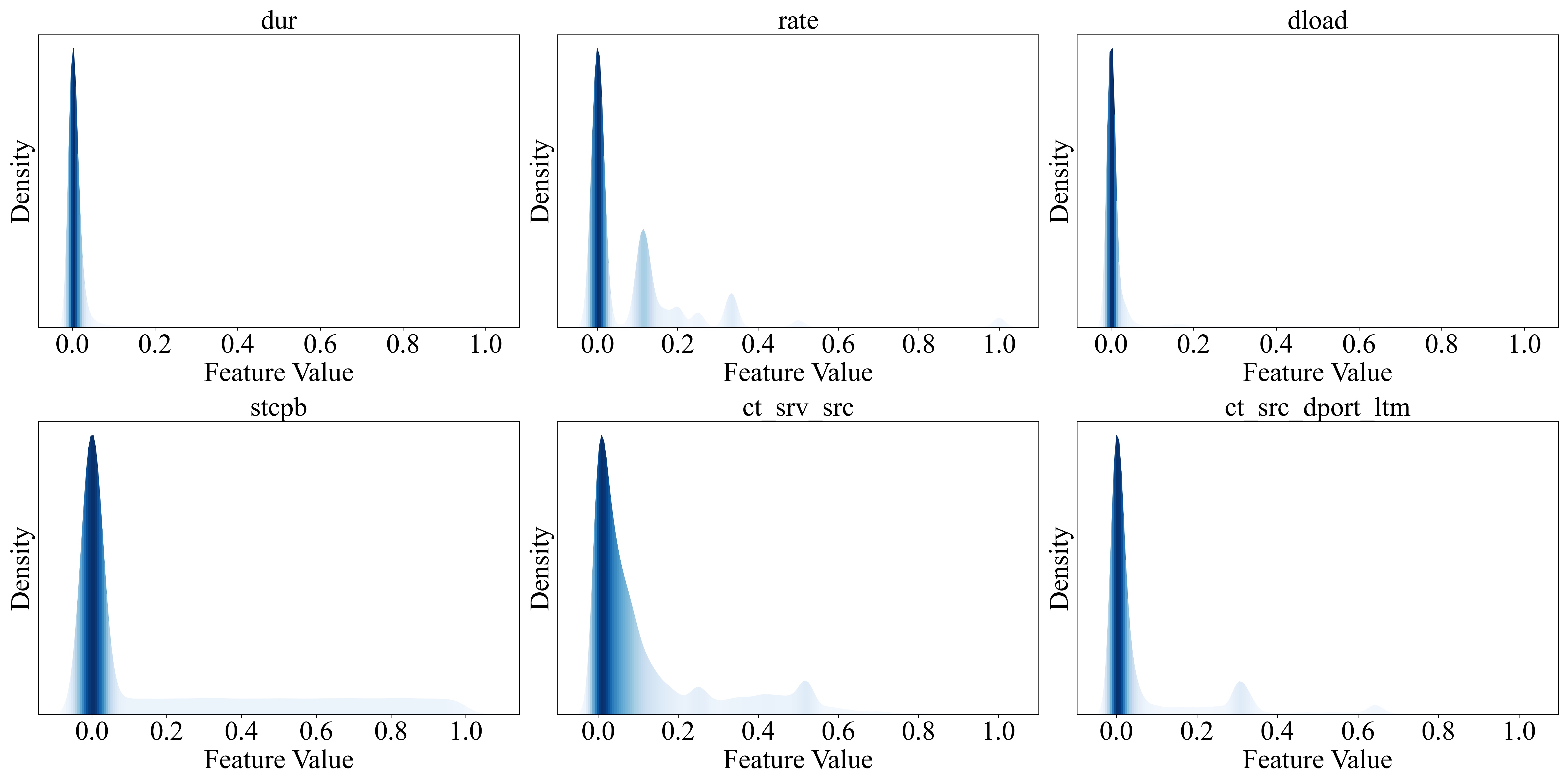}%
\label{fig_1-1b}}
\caption{The density distribution of NSL-KDD and UNSW-NB16 datasets. The x-axis represents the feature value, while the y-axis indicates the density, and darker colors correspond to higher density.}
\label{fig_1-1}
\end{figure}

Feature sparsification can hinder model convergence and restrict gradient propagation \cite{evci2019difficulty, ma2017stabilized, yi2020not}. Deep learning relies on backpropagation to update parameters, but sparse inputs often cause many neurons to remain inactive (i.e. output zero), yielding zero gradients and blocking parameter updates. These “silent” neurons reduce the expressive capacity of the network, limiting its ability to model data distributions effectively. As a result, convergence slows and training may stagnate. Fundamentally, sparsity weakens the information flow and narrows the scope of gradient updates, ultimately affecting overall training performance.

\paragraph{\textbf{Challenge 2}} \textit{Lack of an Efficient and Comprehensive Semantic Embedding.} Most existing traffic detection methods mainly follow two paradigms: unsupervised approaches \cite{fu2023detecting, wang2013internet, zhang2014robust}, such as clustering algorithms that model the underlying structure of the data, and supervised approaches \cite{yang2021conditional, zha2025nids, zha2025dm, zha2024skt} that rely on cross-entropy loss as the optimization objective and require manually labeled data for training. These methods can achieve satisfactory performance under ideal conditions with sufficient labeled data.

However, these methods exhibit significant limitations in capturing the complex semantic structure of network traffic. Specifically, "benign" traffic does not constitute a single semantic category but rather comprises multiple subclasses characterized by diverse behavioral patterns. Relying solely on coarse-grained labeling with a generic "benign" tag neglects the inherent semantic heterogeneity, leading to learned representations that lack fine-grained discriminative power. The same issue applies to anomalous traffic: different attack types exhibit distinct behavioral signatures, and using a unified "anomalous" label fails to capture their intrinsic differences. Furthermore, even when fine-grained labels are available to better express semantic characteristics, the embedding representation must still preserve intraclass diversity to enhance the model’s generalization capacity. Maintaining such structural diversity not only prevents overfitting to specific feature patterns but also improves the robustness and adaptability of the model.

\textbf{\textit{Our work.}} We propose a novel traffic detection method, named \textbf{\textit{FlowXpert}}, to solve the above issues. Firstly, we design a novel feature extraction scheme that constructs contextual feature vectors for model training by associating the destination node of a network flow with its source host node. This approach eliminates many features of the sparse problems commonly found in traditional methods, thus avoiding their negative impact on model convergence. In addition, contextual characteristics demonstrate superior generalization performance, enhancing the practicality of the model in real-world applications. Secondly, we incorporate the unsupervised DBSCAN clustering algorithm \cite{schubert2017dbscan} and employ contrastive learning \cite{chopra2005learning, khosla2020supervised} to train the embedding space, which collectively improves the overall performance of the model in traffic detection, while also improving its generalization ability and adaptability. This process can be performed efficiently on downsampled data. Finally, we conduct experiments on the MAWI real-world dataset \cite{mawi_dataset, fontugne2010mawilab}, avoiding the use of commonly adopted synthetic datasets such as CICIDS-2017 \cite{CICIDS2017} and NSL-KDD \cite{Tavallaee2009}, to better approximate real-world deployment scenarios and validate the effectiveness of \textit{FlowXpert}. We perform a comprehensive evaluation of \textit{FlowXpert} from multiple dimensions, including cross-test, generalization test, encrypted traffic identification, ablation studies, comparative experiments and real-time performance tests, all of which confirm the effectiveness and progress of the method. In addition, we rigorously control the experimental setup to avoid potential label leakage issues often found in previous studies. All experiments are conducted using traffic data collected during two different time periods, thereby ensuring the robustness and credibility of the experimental results.

\textbf{\textit{Contributions.}} This study makes the following contributions:
\begin{itemize}
    \item We propose a novel feature extraction tool that not only incorporates selected traditional flow-related features, but also integrates contextual semantic information associated with the source host. This approach effectively mitigates the issue of sparsity inherent in existing flow features.
    \item We propose an unsupervised traffic embedding method designed to enhance the model’s ability to represent and interpret traffic features, thereby improving the overall performance of intrusion detection. Furthermore, we provide a rigorous theoretical proof of the convergence properties of the proposed embedding approach.
    \item We perform cross-test, ablation studies, and comparative experiments on the MAWI real-world data set to thoroughly test the effectiveness of the proposed \textit{FlowXpert}. The experimental results demonstrate that it achieves outstanding performance in various evaluation metrics, even when handling encrypted traffic.
\end{itemize}

The remainder of the paper is organized as follows. Section II describes the design of \textit{FlowXpert}. Section III presents the details of Feature Extraction, and Section IV introduces the training process of Embedding. In Section V, we experimentally evaluate the performance of our method. Section VI reviews the related work. Finally, we conclude this paper in Section VII.

\section{DESIGN OF FLOWXPERT}
In this section, we present the overall architecture of \textbf{\textit{FlowXpert}}, as shown in Fig. 2. \textit{FlowXpert} consists of three important components: \textbf{\textit{Feature Extraction}} and \textbf{\textit{Embedding}}, and \textbf{\textit{Model Training}}.
\begin{figure*}[htbp]
\centering
\includegraphics[width=0.8\textwidth]{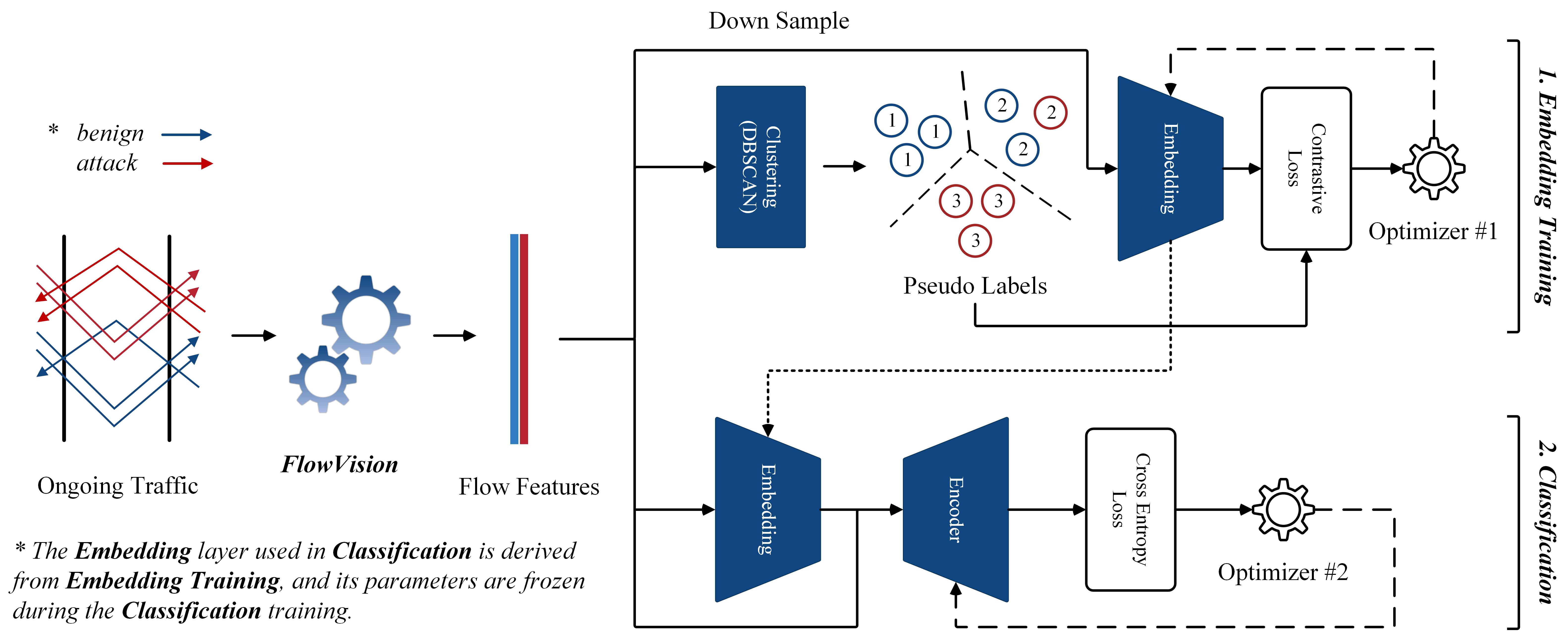}
\caption{The framework of \textbf{\textit{FlowXpert}}.}
\label{fig_1-1}
\end{figure*}

\paragraph{\textbf{Feature Extraction}} As highlighted in \textbf{\textit{Challenge 1}}, the widely used flow-based features impede both model convergence and detection accuracy. To improve generalization, we propose \textbf{\textit{FlowVision}}, a context-aware feature extraction tool. \textit{FlowVision} discards all highly sparse flow features, retaining only nine essential flow attributes and supplementing them with four context-related host features. Detailed specifications are provided in Section III. Furthermore, discrete features are processed with one-hot encoding, and continuous features undergo min–max normalization.

\paragraph{\textbf{Embedding}} As discussed in \textbf{\textit{Challenge 2}}, using supervised cross-entropy loss as the optimization objective tends to cause overfitting and fails to efficiently and comprehensively capture the semantic structure of network traffic. To address this, we propose an embedding training method for traffic detection that operates effectively with only down-sampled data. This method leverages pseudo-labels generated via the DBSCAN clustering algorithm \cite{schubert2017dbscan} and incorporates contrastive learning \cite{chopra2005learning} to enhance semantic representation. Detailed methodology is presented in Section IV.

\paragraph{\textbf{Model Training}} After the \textit{Embedding} model is trained, its parameters are frozen and directly utilized in the subsequent intrusion detection model training. The semantic vectors $V_E$ generated by the \textit{Embedding} module are concatenated with the original input vectors $X$ to form a fused representation with a residual structure. This fused vector $X_\text{Concat}$ is then fed into the encoder module. Finally, the intrusion detection model is trained using cross-entropy loss as the optimization objective. The overall process is illustrated as follows:
\begin{equation}
\begin{aligned}
    X_E &= \text{Embedding}(X), \\
    X_\text{Concat} &= \text{Concat}(X, X_E), \\
    Logits &= \text{Linear}(\text{Enc}(X_\text{Concat})), \\
    L_{ce} &= \text{CELoss}(Logits, Y),
\end{aligned}
\end{equation}
where $\text{Linear}(\cdot)$, $\text{Enc}(\cdot)$ and $\text{CELoss}(\cdot)$ denote fully-connected layer, encoder and the cross entropy loss.

\section{Feature Extraction}
In this section, we systematically analyze the limitations of widely used flow features in public data sets to optimize traffic detection models \cite{CICIDS2017, CICIDS2018}. Consequently, we propose a context-aware alternative feature extraction framework that enhances feature representation and improves model generalization.

\subsection{Impact of Sparse Features on Model Optimization}
The presence of numerous time- and length-related features often results in sparse input flow vectors. To better characterize their impact on model optimization, we provide the following two definitions.

\begin{mdframed}[backgroundcolor=lightgray, linecolor=lightgray, linewidth=0pt]
\textbf{\textit{Definition 1.}} \textit{Let the input vector $X=[x_1, x_2, ..., x_d]^T$ denote a feature vector, where $d$ is the dimension. The input is called sparse if the majority of the components of $x$ are close to zero, i.e.,}
\begin{equation}
   \forall i \in \{1, 2, ..., d\}, |x_i| \ll 1 ~\text{for most}~i.
\end{equation}
\end{mdframed}

\begin{mdframed}[backgroundcolor=lightgray, linecolor=lightgray, linewidth=0pt]
\textbf{\textit{Definition 2.}} \textit{Consider a neural network layer where the output $a$ is computed as the activation of the weighted sum of the input vector $X$, i.e.,}
\begin{equation}
    z = \text{W}^T X + \text{b}, a=\Phi(z),
\end{equation}
\textit{where $\text{W}$ is the weight vector, $b$ is the bias term, and $\Phi(\cdot)$ is an activation function (e.g., ReLU or Sigmoid). }
\end{mdframed}

Under the settings of \textbf{\textit{Definition 1}} and \textbf{\textit{Definition 2}}, the following theorem holds.

\begin{mdframed}[backgroundcolor=lightgray, linecolor=lightgray, linewidth=0pt]
\textbf{\textit{Theorem 1.}} \textit{Let $X$ be a sparse vector, then the norm of the gradient $\left \| \frac{\partial L}{\partial \text{W}} \right \|$ is bounded above by a small constant, leading to a slower convergence rate of the optimization algorithm.}
\end{mdframed}

\textbf{\textit{Corollary 1.}} Consider the gradient of the loss function $L(a,y)$ with respect to the weights: 
\begin{equation}
    \frac{\partial L}{\partial \text{W}} = \frac{\partial L}{\partial a} \cdot \frac{\partial a}{\partial z} \cdot \frac{\partial z}{\partial \text{W}} = \delta \cdot X.
\end{equation}

If $X$ is sparse (by \textbf{\textit{Definition 1}}), most elements of $X$ are close to zero, so the gradient:
\begin{equation}
    \frac{\partial L}{\partial \text{W}} \approx 0.
\end{equation}

This leads to a small update for the weights during training. As a result, the convergence rate of the model is slowed down because the parameter updates are minimal. In deep neural networks, the gradient at each layer depends on the gradient propagated from the previous layer. When the input is sparse, the gradients in each layer will be small. Specifically, the gradient at the k-th layer is computed as follows:
\begin{equation}
    \frac{\partial L}{\partial \text{W}^{(k)}} = \delta^{(k)} \cdot X^{(k)},
\end{equation}
where $\delta^{(k)}$ is the error term for the k-th layer and $X^{(k)}$ is the input to the k-th layer. If $X^{(k)}$ is sparse, the gradient $\frac{\partial L}{\partial \text{W}^{(k)}}$ will be small, and this effect will propagate through all layers, causing the gradient to disappear.

Therefore, during gradient descent optimization, the step size for weight updates is also small:
\begin{equation}
    \text{W}^{(t+1)} = \text{W}^{(t)} - \eta \cdot \frac{\partial L}{\partial \text{W}},
\end{equation}
where $\eta$ is the learning rate. Since $\frac{\partial L}{\partial \text{W}}$ is small, the weight update step becomes smaller, leading to slower convergence.

\subsection{FlowVision}
As discussed in Section III.A, the widely adopted flow-based features commonly used in existing approaches have shown significant adverse effects on model convergence in practical scenarios. Based on this observation, we discard most highly sparse flow features and propose a novel feature extraction tool, named \textit{FlowVision}. Unlike existing tools such as CICMeterFlow, \textit{FlowVision} not only retains a subset of fundamental statistical flow features, but also places greater emphasis on contextual semantic information of flows. The overall data structure of \textit{FlowVision} is illustrated in Fig. 3. Specifically, \textit{FlowVision} uses five tuples as keys to segment and aggregate the pcap packets into distinct flows. Additionally, each flow node is enriched with contextual semantic attributes related to its source IP, forming a lightweight semantic structure that is more concise and efficient compared to directly constructing a full graph structure.
\begin{figure}[htbp]
\centering
\includegraphics[width=0.45\textwidth]{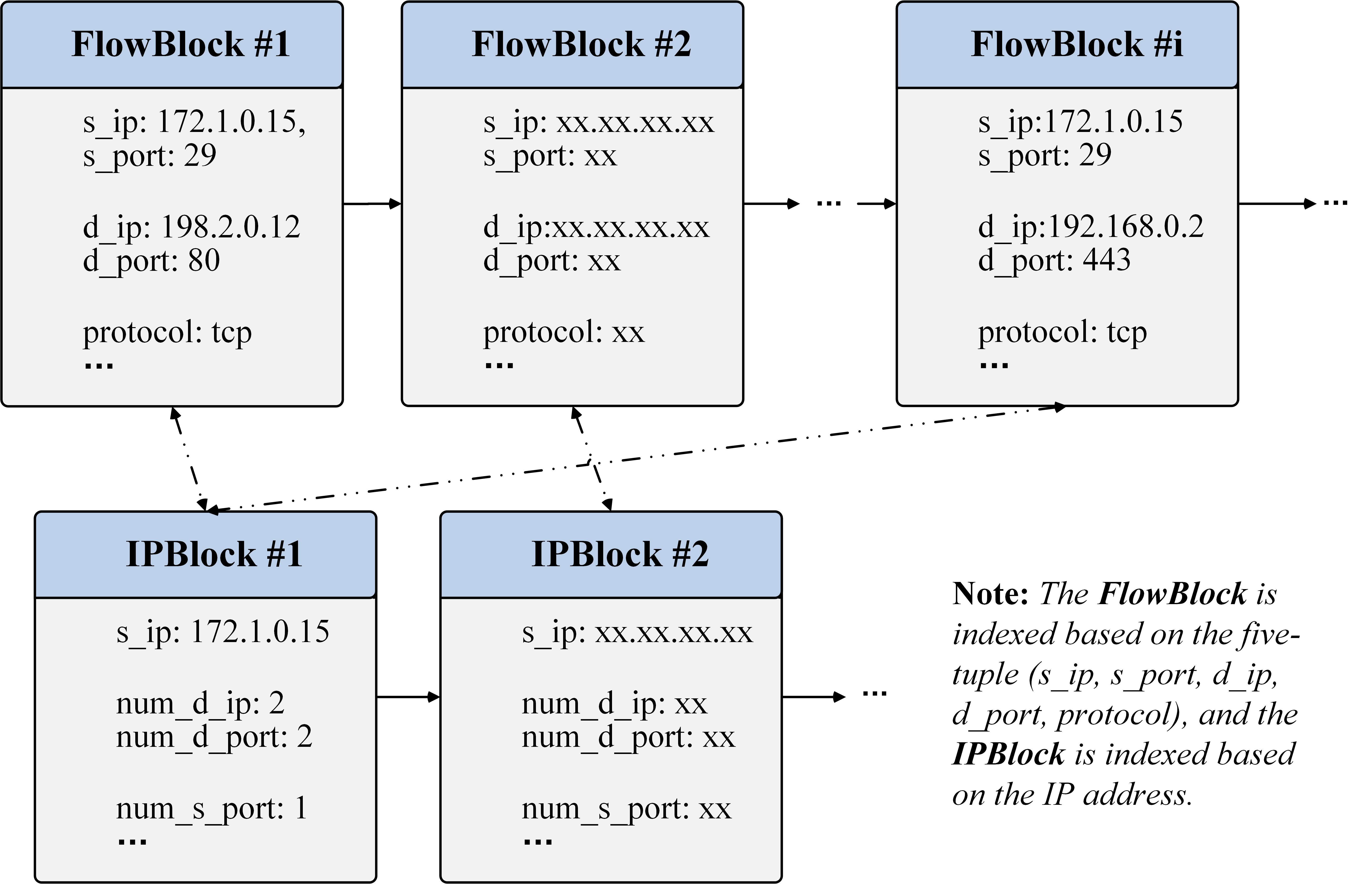}
\caption{The data structure of \textbf{\textit{FlowVision}}.}
\label{fig_3-1}
\end{figure}

\textit{FlowVision} extracts a total of 13 key features for model training and intrusion detection, encompassing both flow-level and contextual semantic information, as detailed in Table I. The flow-level features include protocol type, flow duration, inter-arrival time (IAT), the number of SYN/RST/FIN packets, and packet rate. Contextual features capture higher-level semantics, such as the total number of source ports initiated by a given source IP, the number of destination IPs and ports accessed, and the rate at which the source IP establishes connections over time.
\begin{table}[htbp]
\centering
\caption{FEATURES USED IN FLOWVISION}
\begin{tabularx}{0.49\textwidth}{
    >{\centering\arraybackslash}p{0.09\textwidth}|
    >{\centering\arraybackslash}p{0.08\textwidth}|
    >{\centering\arraybackslash}X
    }
\toprule
\textbf{Type} & \textbf{Features} & \textbf{Description} \\

\midrule
\multirow{11}{*}{Flow-Level} & protocol & The protocol of transmission layer. \\
& flow\_dur & The duration of the flow. \\
& \multirow{2}{*}{iat\_mean} & The mean value of inter-arrival time (iat) between packets. \\
& \multirow{2}{*}{iat\_std} & The standard deviation of iat between packets.\\
& fin\_num & The total FIN packets of the flow. \\
& syn\_num & The total SYN packets of the flow. \\
& rst\_num & The total RST packets of the flow. \\
& pkt\_num & The total packets of the flow. \\
& pkts\_per\_sec & The packets per second of the flow. \\ 
\midrule
\multirow{8}{*}{Contextual} & \multirow{2}{*}{num\_s\_port} & The number of distinct source ports used by the same source IP. \\
& \multirow{2}{*}{num\_d\_ip} & The number of distinct destination IPs contacted by the same source IP. \\
& \multirow{2}{*}{num\_d\_port} & The number of distinct destination ports accessed by the same source IP. \\
& \multirow{2}{*}{con\_per\_sec} & The number of connections established per second by the source IP. \\
\bottomrule
\end{tabularx}
\label{tab:3-1}
\end{table}

\section{Embedding Training}
To comprehensively and efficiently capture the semantic information of traffic characteristics, we propose an unsupervised \textit{Embedding} training method, mainly composed of a \textit{DBSCAN} clustering component \cite{schubert2017dbscan} and a contrastive learning strategy \cite{chopra2005learning, khosla2020supervised}. 

\subsection{Clustering (DBSCAN)}
As described in \textbf{\textit{Challenge 2}}, due to the nature of traffic detection tasks, a single label often fails to accurately and comprehensively represent all subtypes within a given traffic class. In other words, traffic samples belonging to the same class may exhibit internal structural differences, necessitating finer-grained labeling to distinguish between subtypes. This objective can be effectively achieved using unsupervised algorithms, such as clustering methods \cite{pan2025towards, schubert2017dbscan, zhang2025structured}.

Common clustering algorithms include KMeans \cite{zhang2025structured} and DBSCAN \cite{schubert2017dbscan}. KMeans is a distance-based method that requires the number of clusters to be specified in advance and assigns each sample deterministically to a cluster. In contrast, DBSCAN does not require a predefined number of clusters, can identify clusters of arbitrary shape, and allows certain samples to remain unclustered, and these are treated as noise or outliers. During the embedding learning process, it is inadvisable to assign fine-grained pseudolabels to all samples indiscriminately, especially for those with atypical characteristics. Aggressive pseudo-labeling of such samples may lead to inaccurate representation. The clustering mechanism of DBSCAN is well-suited to this requirement, as it naturally accommodates noise points. In addition, extensive prior research has demonstrated that density-based clustering methods, such as DBSCAN, often yield superior performance in traffic detection tasks. This procedure is formulated as follows:
\begin{equation}
    P = DBSCAN(X),
\end{equation}
where $P$ denotes the pseudo labels correspond to $X$.

\subsection{Embedding Layer}
To obtain more fine-grained semantic representations of network traffic data, we leverage the pseudolabels generated by Eq. (8) to train the embedding layer. Let $X_i$ and $X_j$ denote two traffic samples. Based on the pseudo-labels derived from Eq. (8), a new binary label $B$ can be constructed, $B=0$ if the sample $X_i$ and $X_j$ belong to the sample cluster and $B=1$ otherwise. Furthermore, let $E(X_E^i, X_E^j)$ indicate the Euclidean distance between the corresponding semantic vectors $X_E^i$ and $X_E^j$ of the two samples in the embedding space. This can be formulated as follows:
\begin{equation}
    E(X_E^i, X_E^j) = ||X_E^i - X_E^j||.
\end{equation}

We aim to ensure that the following condition is satisfied and that the embedding layer training is conducted accordingly.
\begin{mdframed}[backgroundcolor=lightgray, linecolor=lightgray, linewidth=0pt]
\textbf{\textit{Condition 1.}} \textit{$\exists~m>0$, such that $E_{pos} + m < E_{neg}$, where $E_{pos}=E(X_E^i, X_E^j)$, where $B = 0$; and $E_{neg}=E(X_E^i, X_E^j)$, where $B=1$.}
\end{mdframed}

\subsection{Contrastive Loss}
Inspired by \cite{chopra2005learning}, we introduce contrastive learning to implement the aforementioned idea. Assuming that the loss function depends only on the input $X$, the weight matrix $\text{W}_E$ of the embedding layer, and the binary label $B$, the loss function can be formulated as follows:
\begin{equation}
\begin{aligned}
    L(\text{W}_E) &= \sum_{i,j}^N L(\text{W}_E, B, (X_i, X_j)), \\
    L(\text{W}_E, B, (X_i, X_j)) &= (1-B) \cdot L_{pos}(E(X_i, X_j)) \\
    &+ B \cdot L_{neg}(E(X_i, X_j)).
\end{aligned}
\end{equation}

Then, the total contrastive loss function can be defined as follows:
\begin{equation}
    H(E_{pos}, E_{neg}) = L_{pos}(\text{W}_E,E_{pos}) + L_{neg}(\text{W}_E, E_{neg}).
\end{equation}

Assume that $H(E_{pos}, E_{neg})$ is convex in its two arguments (do not assume that it is convex with respect to $\text{W}_E$).

Due to the use of the gradient descent algorithm, the following condition holds.
\begin{mdframed}[backgroundcolor=lightgray, linecolor=lightgray, linewidth=0pt]
\textbf{\textit{Condition 2.}} \textit{The negative gradient of $H(E_{pos}, E_{neg})$ on the margin line $E_{pos} + m = E_{neg}$ has a positive dot product with direction $\left[ -1, 1\right]$.}
\end{mdframed}

To formalize the above idea and clarify our reasoning, the following theorem and its proof are presented.

\begin{mdframed}[backgroundcolor=lightgray, linecolor=lightgray, linewidth=0pt]
\textbf{\textit{Theorem 2.}} \textit{Let $H(E_{pos}, E_{neg})$ have its minimum at infinity and assume that there exists a weight vector $\text{w}$ satisfying \textbf{Condition 1}. If \textbf{Condition 2} holds, then the minimization of $H(E_{pos}, E_{neg})$ with respect to $\text{w}$ will produce a solution $\text{w} $ that satisfies \textbf{Condition 1}.}
\end{mdframed}

\textbf{\textit{Proof.}} Let $E_{pos}^*$ denote the data point located on the margin line $E_{pos} + m = E_{neg}$, for which $H(E_{pos}, E_{neg})$ is minimum, that is,
\begin{equation}
    E_{pos}^* = \text{argmin} \{H(E_{pos}, E_{pos} + m)\}.
\end{equation}

Since \textbf{\textit{Condition 2}} holds and the $H(E_{pos}, E_{neg})$ is convex, it follows that:
\begin{equation}
    H(E_{pos}^*, E_{pos}^* + m) \leq H(E_{pos}, E_{neg}),
\end{equation}
when $E_{pos} + m > E_{neg}$.

Consider a data point at a distance $\epsilon$ from $(E_{pos}^*, E_{pos}^* + m)$ and satisfying $E_{pos} + m < E_{neg}$, that is,
\begin{equation}
    (E_{pos}^* - \epsilon, E_{pos}^* + \epsilon + m).
\end{equation}

By first-order Taylor expansion, it follows that:
\begin{equation}
\begin{aligned}
    &~~~~H(E_{pos}^* - \epsilon, E_{pos}^* + \epsilon + m) \\
    &= H(E_{pos}^*, E_{pos}^* + m) - \epsilon \frac{\partial H}{\partial E_{pos}} + \epsilon \frac{\partial H}{E_{neg}} + O(\epsilon^2) \\
    &= H(E_{pos}^*, E_{pos}^* + m) + \epsilon \left[\frac{\partial H}{\partial E_{pos}}, \frac{\partial H}{E_{neg}} \right] \begin{bmatrix} -1 \\ 1 \end{bmatrix} + O(\epsilon^2). \\
\end{aligned}
\end{equation}

By \textbf{\textit{Condition 2}}, it follows that:
\begin{equation}
    \left[\frac{\partial H}{\partial E_{pos}}, \frac{\partial H}{E_{neg}} \right] \begin{bmatrix} -1 \\ 1 \end{bmatrix} < 0.
\end{equation}

For sufficiently small $\epsilon$ (we set $\epsilon = m$ in this paper), 
\begin{equation}
    H(E_{pos}^* - \epsilon, E_{pos}^* + \epsilon + m) \leq H(E_{pos}^*, E_{pos}^* + m).
\end{equation}

Thus, there exists a data point in the region where $E_{pos} + m < E_{neg}$ such that the loss function value is always lower than that of any data point in the region where $E_{pos} + m > E_{neg}$. This completes the proof.

\subsection{Practical Application}
Building on Eq. (13) and Eq. (17), one can show that there exists a sufficiently small $\epsilon$ such that minimizing the loss function in Equation (11) yields a weight matrix $\text{W}_E$ satisfying \textbf{\textit{Condition 1}}, thus guaranteeing convergence. When $\epsilon$ is sufficiently small, the minimization of the objective $H$ drives $E_{pos}^* \to 0$; that is, the semantic distance between samples sharing the same pseudo-label approaches zero, while the distance between samples with different pseudo-labels exceeds the margin $m$. During embedding training, however, our goal is not to force intra-label distances to vanish. Instead, we keep these distances within a reasonable threshold to preserve semantic diversity and, consequently, enhance model generalization. In this study, we set $\epsilon=m$. Under this choice, the contrastive learning strategy for positive and negative sample pairs is formulated as follows:
\begin{itemize}
    \item If $X_E^i,~X_E^j$ belong to the same pseudo label, then $||X_E^i - X_E^j|| \leq m$;
    \item If $X_E^i,~X_E^j$ not belong to the same pseudo label, then $||X_E^i - X_E^j|| \ge 2m$.
\end{itemize}

More specifically, the loss function $L_{contrastive}$ is defined as follows:
\begin{equation}
\begin{aligned}
\mathcal{L}_{\text{Contrastive}} & = \frac{\sum_{i,j} \mathbb{I}(B_i = B_j) \cdot \max(0, E(X_E^i, X_E^j) - m)}{\sum_{i,j} \mathbb{I}(B_i = B_j)} \\
& + \frac{\sum_{i,j} \mathbb{I}(B_i \neq B_j) \cdot \max(0, 2m - E(X_E^i, X_E^j))}{\sum_{i,j} \mathbb{I}(B_i \neq B_j)},
\end{aligned}
\end{equation}
where $\mathbb{I}(\cdot) == 1$ if condition $\cdot$ is true.

To improve training efficiency, the embedding model is trained using a subset of downsampled data from the entire data set, which represents approximately 2\% of the total data. The embedding architecture consists of multiple fully connected layers, with details provided as follows.
\begin{align}
Embedding: 
\begin{cases}
h_1 = \text{Linear}_{|X| \rightarrow 128}(X), \\
h_1^B = \text{BatchNorm}(h_1), \\
h_1^R = \text{LeakyReLU}(h_1^B), \\
X_E = \text{Linear}_{128 \rightarrow 16}(h_1^R), \\
\end{cases}
\end{align}

\begin{align}
Encoder: 
\begin{cases}
X_{Concat} = \text{Concat}(X,X_E), \\
h_1 = \text{Linear}_{|X| + 16 \rightarrow 512}(X_{Concat}), \\
h_1^R = \text{LeakyReLU}(h_1), \\
h_2 = \text{Linear}_{512 \rightarrow 256}(h_1^R), \\
h_2^R = \text{LeakyReLU}(h_2), \\
X_{Enc} = \text{Linear}_{256 \rightarrow 128}(h_2^R), \\
\end{cases}
\end{align}

\begin{equation}
Classifier:~Logits = \text{Linear}_{128 \rightarrow 2} (X_{Enc}), \\
\end{equation}
where $\text{Linear}(\cdot)$, $\text{BatchNorm}(\cdot)$, $\text{LeakyReLU}(\cdot)$ and $\text{Concat}(\cdot)$ denote fully connected layer, batch normalization, activation function, and concatenated operation.

\section{Experimental Evaluation}
\subsection{Experiment Setup}
\paragraph{\textbf{Implementation}} Our prototype is implemented in Python (3.8.12) and C++ (g++ 10.5.0), the \textit{FlowVision} is built using C++, and the other modules are built using scikit-learn \cite{pedregosa2011scikit} and PyTorch (2.0.1+cu117) \cite{paszke2019pytorch}. 

\paragraph{\textbf{Datasets}} All experiments in this study were conducted on the real-world and publicly available MAWI dataset \cite{fontugne2010mawilab, mawi_dataset}, rather than using simulated datasets such as NSL-KDD \cite{Tavallaee2009} or CICIDS-2017 \cite{CICIDS2017}, in order to more accurately evaluate the practical applicability of \textit{FlowXpert} in real-world scenarios. The MAWI data set provides raw traffic data in PCAP format, encompassing a wide range of attack types, including DoS, port scanning, flooding attacks, and brute-force attacks, as well as imbalanced distributions of benign and malicious traffic. It also includes various network protocols such as HTTP, SMTP, FTP, SSH, and encrypted traffic such as HTTPS. For our experiments, we selected traffic data from two different time periods. March 2021 and June 2023. Specifically, data from March 1, 2021, and June 3, 2023, were used for model training with a 3-fold cross-test (2:1 split), while data from March 8, 15, and 22 of 2021, and June 10, 18, and 25 of 2023 (each approximately one week apart) were used to evaluate the generalization performance of the model.

To ensure the applicability of the proposed method in real-world scenarios, both the data set and the evaluation methodology were designed to meet the following conditions.
\begin{itemize}
    \item The experiments were carried out using the complete MAWI data set without any data removal. Although sampling was applied during the embedding training phase, classification training and generalization evaluation were performed on the full data set.
    \item Features such as IPs and ports, which can potentially reveal label information, were excluded from the training. In addition, potential associated cheating behaviors were explicitly avoided. For example, most samples in the data set are concentrated on a limited set of IP or ports.
    \item Due to the inherent class imbalance between benign and malicious traffic, a metric such as accuracy was not used. Instead, detection performance metrics were separately evaluated for benign and malicious traffic to provide a more comprehensive and fair assessment.
\end{itemize}

\paragraph{\textbf{Baselines}} We selected three methods for comparative experiments to demonstrate the correctness and improvement effect of our method, the details are as follows:
\begin{itemize}
    \item \textit{Kitsune} \cite{mirsky2018kitsune}. Mirsky et al. proposed an unsupervised network intrusion detection method centered on KitNET, which uses an ensemble of lightweight autoencoders to detect anomalies through reconstruction error analysis of network traffic features.
    \item \textit{CVAE-EVT} \cite{yang2021conditional}. Yang et al. proposed a two-stage learning method that combines conditional variational autoencoders and extreme value theory to build a hierarchical intrusion detection system. 
    \item \textit{HyperVision} \cite{fu2023detecting}. Fu et al. proposed using an in-memory flow interaction graph with unsupervised graph learning to detect encrypted malicious traffic.
\end{itemize}

All comparative methods were evaluated using the same source data and official implementations from their GitHub repositories. Only minor modifications were made, primarily for compatibility purposes such as adapting data loading interfaces. For \textit{HyperVision}, processing datasets with many unique IP addresses led to excessive memory usage, exhausting a 256 GB DRAM server. Consequently, we adopted data sampling to complete the experiments.

\paragraph{\textbf{Metrics}} We use five metrics to evaluate the performance of \textit{FlowXpert}, including three commonly used metrics in machine learning algorithms: \textit{Precision}, \textit{Recall}, and \textit{F1-Score} \cite{japkowicz2011evaluating}. Furthermore, we include two metrics that are crucial for practical deployment: \textit{Latency} and \textit{Throughput}.

\subsection{Cross Evaluation of Traffic Classification}
To comprehensively evaluate the detection performance of \textit{FlowXpert} in real-world network environments, we performed 3-fold cross-test experiments on the MAWI data set. Specifically, the original data set was randomly divided into three mutually exclusive subsets. In each iteration, two subsets were used for training while the remaining subset served as the test set. This process was repeated three times to ensure the stability and generalizability of the model on different data partitions. The results are presented in Table II.
\begin{table}[htbp]
    \centering
    \caption{CROSS EVALUATION ON THE MAWI DATASET}
    \begin{tabularx}{0.48\textwidth}{
        >{\centering\arraybackslash}p{0.07\textwidth} |
        >{\centering\arraybackslash}p{0.04\textwidth} |
        >{\centering\arraybackslash}p{0.07\textwidth} |
        >{\centering\arraybackslash}X 
        >{\centering\arraybackslash}X 
        >{\centering\arraybackslash}X 
        }
        \toprule
        \textbf{Date} & \textbf{Fold} & \textbf{Label} & \textit{Pre} & \textit{Rec} & \textit{F1} \\
        \midrule
        \multirow{7}{*}{2021/3/1} & \multirow{2}{*}{1} & Benign & 98.98\% & 99.85\% & 99.41\% \\
        & & Malicious & 99.12\% & 94.34\% & 96.67\% \\
        \cmidrule{2-6}
        & \multirow{2}{*}{2} & Benign & 98.97\% & 99.76\% & 99.36\% \\
        & & Malicious & 98.60\% & 94.34\% & 96.42\% \\
        \cmidrule{2-6}
        & \multirow{2}{*}{3} & Benign & 98.94\% & 99.87\% & 99.40\% \\
        & & Malicious & 99.23\% & 94.14\% & 96.62\% \\

        \midrule
        \multirow{7}{*}{2023/6/3} & \multirow{2}{*}{3} & Benign & 99.26\% & 99.32\% & 99.29\% \\
        & & Malicious & 96.40\% & 96.07\% & 96.23\% \\
        \cmidrule{2-6}
        & \multirow{2}{*}{2} & Benign & 99.24\% & 99.24\% & 99.24\% \\
        & & Malicious & 95.98\% & 95.97\% & 95.98\% \\
        \cmidrule{2-6}
        & \multirow{2}{*}{1} & Benign & 99.20\% & 99.36\% & 99.28\% \\
        & & Malicious & 96.57\% & 95.75\% & 96.16\% \\
        \bottomrule
    \end{tabularx}
    \label{tab:table5-1}
\end{table}

As shown in the three sets of experimental results for March 1, 2021, \textit{FlowXpert} achieved a recall rate that exceeded 99.5\% for benign traffic and a recall rate consistently higher than 94\% for malicious traffic, indicating the model's strong capability in identifying both types of traffic. In terms of precision, both Benign and Malicious traffic achieved values close to 99\%, further demonstrating the effectiveness of the model in controlling false positives. Regarding the F1 score, the model achieved approximately 99.5\% for benign traffic and around 96.5\% for malicious traffic, reflecting a high level of overall detection performance, with a good balance between precision and recall. In particular, the false positive rate for benign traffic was kept below 0.5\%, while the detection performance for malicious traffic remained robust, validating the practicality and reliability of the model in real-world traffic scenarios. Similarly, in the three experiments conducted on June 3, 2023, \textit{FlowXpert} maintained a recall rate above 99\% for Benign traffic, with precision and F1 scores reaching approximately 99\% and 96\%, respectively, closely aligned with the results of 2021. This further confirms the stability of the model and its ability to generalize across different time periods. Taken together, these results demonstrate that \textit{FlowXpert} not only effectively distinguishes between Benign and Malicious traffic, but also maintains consistent and high detection performance in network traffic data collected at different times.

\subsection{Evaluation of Ablation Experiments}
To further validate the effectiveness of the embedding layer in enhancing model detection performance, we designed and conducted an ablation study using the data set from March 1, 2021. Specifically, we constructed a baseline model, referred to as \textit{Enc-Cls}, by removing the embedding layer and residual structure from the original architecture, retaining only the encoder module and the subsequent fully connected classifier. Keeping in mind that the input data remained consistent, we compared the detection performance of the two models on the same data set to assess the contribution of the embedding layer. The experimental results are presented in Table III.
\begin{table}[htbp]
    \centering
    \caption{EVALUATION OF ABLATION EXPERIMENTS ON THE MAWI DATASET}
    \begin{tabularx}{0.48\textwidth}{
        >{\centering\arraybackslash}p{0.09\textwidth} |
        >{\centering\arraybackslash}p{0.07\textwidth} |
        >{\centering\arraybackslash}X 
        >{\centering\arraybackslash}X 
        >{\centering\arraybackslash}X 
        }
        \toprule
        \textbf{Combination} & \textbf{Label} & \textit{Pre} & \textit{Rec} & \textit{F1} \\
        \midrule
        \multirow{2}{*}{\textit{Enc + Cls}} & Benign & 98.24\% & 99.55\% & 98.89\% \\
        & Malicious & 97.32\% & 90.22\% & 93.63\% \\
        \midrule
        \multirow{2}{*}{\textbf{\textit{FlowXpert}}} & Benign & 98.98\% & 99.85\% & 99.41\% \\
        & Malicious & 99.12\% & 94.34\% & 96.67\% \\
        \bottomrule
    \end{tabularx}
    \label{tab:table5-2}
\end{table}

The results indicate that even without the embedding layer, the model maintains strong detection performance for benign traffic, with the recall, precision, and F1 score showing no significant deviation from those of the original model. This suggests that for benign traffic—where sample sizes are large and feature distributions are relatively concentrated—the model can still perform well on the classification task without the benefit of feature enhancement from the embedding layer. However, for malicious traffic, the introduction of the embedding layer leads to a substantial improvement in detection performance. Specifically, the recall increases from 90.22\% to 94.34\%, the precision increases from 97.32\% to 99.12\%, and the F1 score improves from 93.63\% to 96.67\%. This performance gain can be mainly attributed to the contrastive learning mechanism embedded within the embedding layer. By enhancing inter-class separability and intra-class consistency, this mechanism enables the model to more effectively capture deep discriminative features within malicious traffic. As a result, it significantly improves the model’s ability to detect complex attack behaviors, particularly in the context of minority-class samples.

\subsection{Visualization of Embedded Representation}
t-SNE visualization provides an intuitive representation of the characteristics of the sample distribution in the vector space, facilitating the exploration of clustering patterns, relative distances and latent structural relationships between different classes. In this study, we used t-SNE to perform a comparative analysis of the Embedding layer feature representations during the contrastive learning training process. Specifically, we recorded the feature embeddings of the original input data as well as those output by the Embedding layer after the 1st, 50th, and 150th training epochs. The experimental results are illustrated in Fig. 4.
\begin{figure*}[htbp]
\centering
\subfloat[Original data.]{\includegraphics[width=0.24\textwidth]{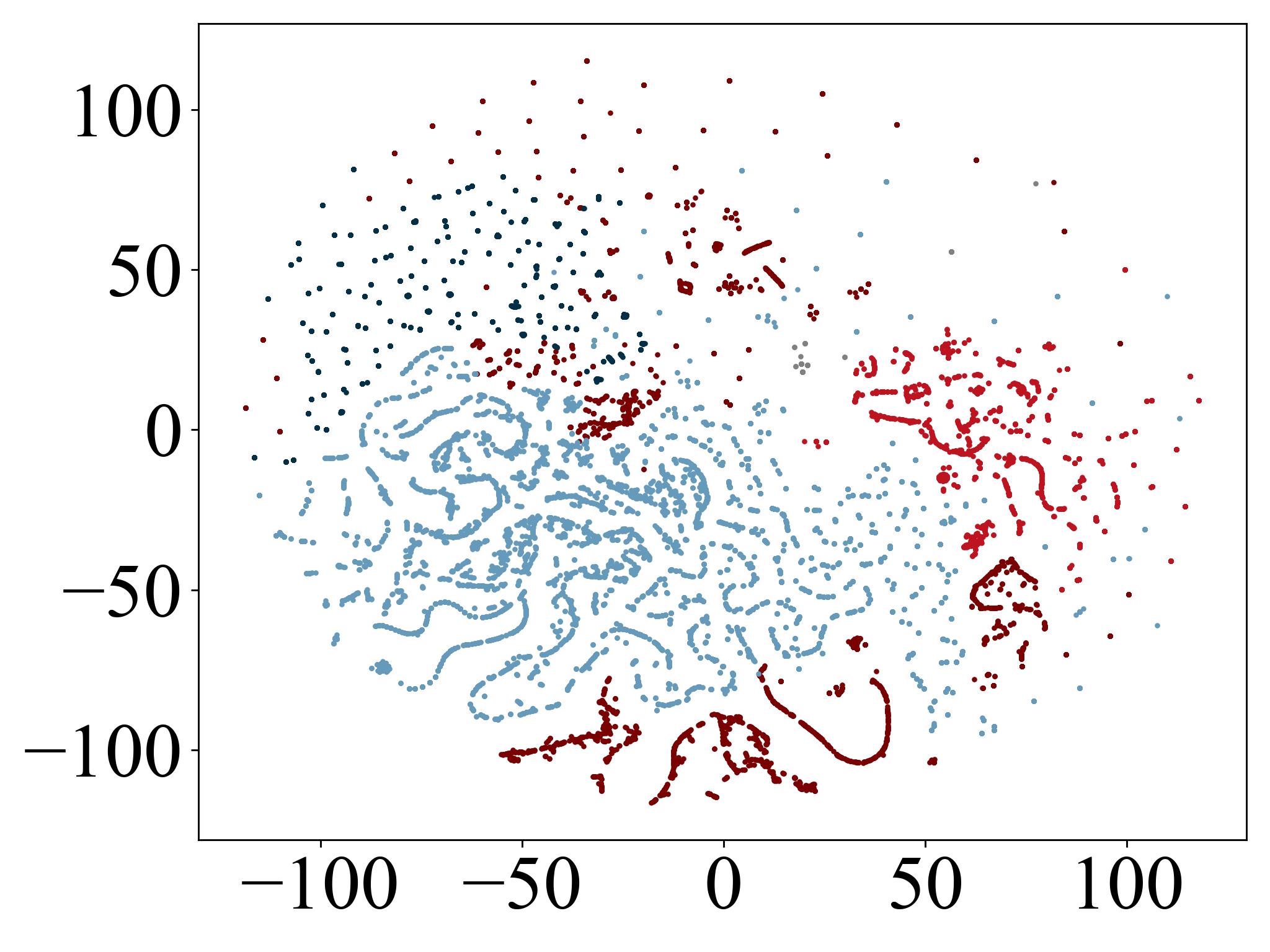}
\label{fig_5-1a}}
\subfloat[Epoch 1.]{\includegraphics[width=0.24\textwidth]{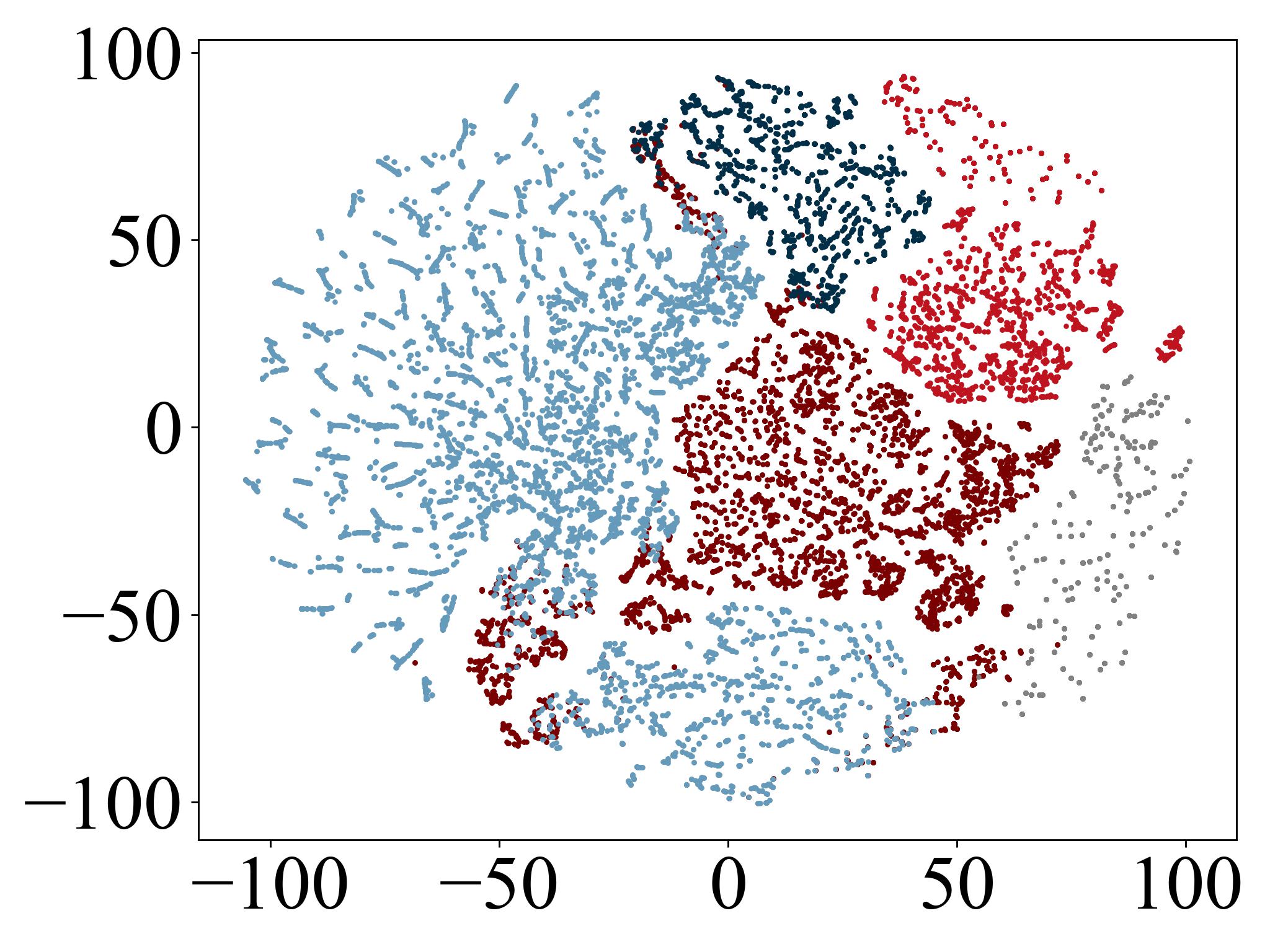}
\label{fig_5-1b}}
\subfloat[Epoch 50.]{\includegraphics[width=0.24\textwidth]{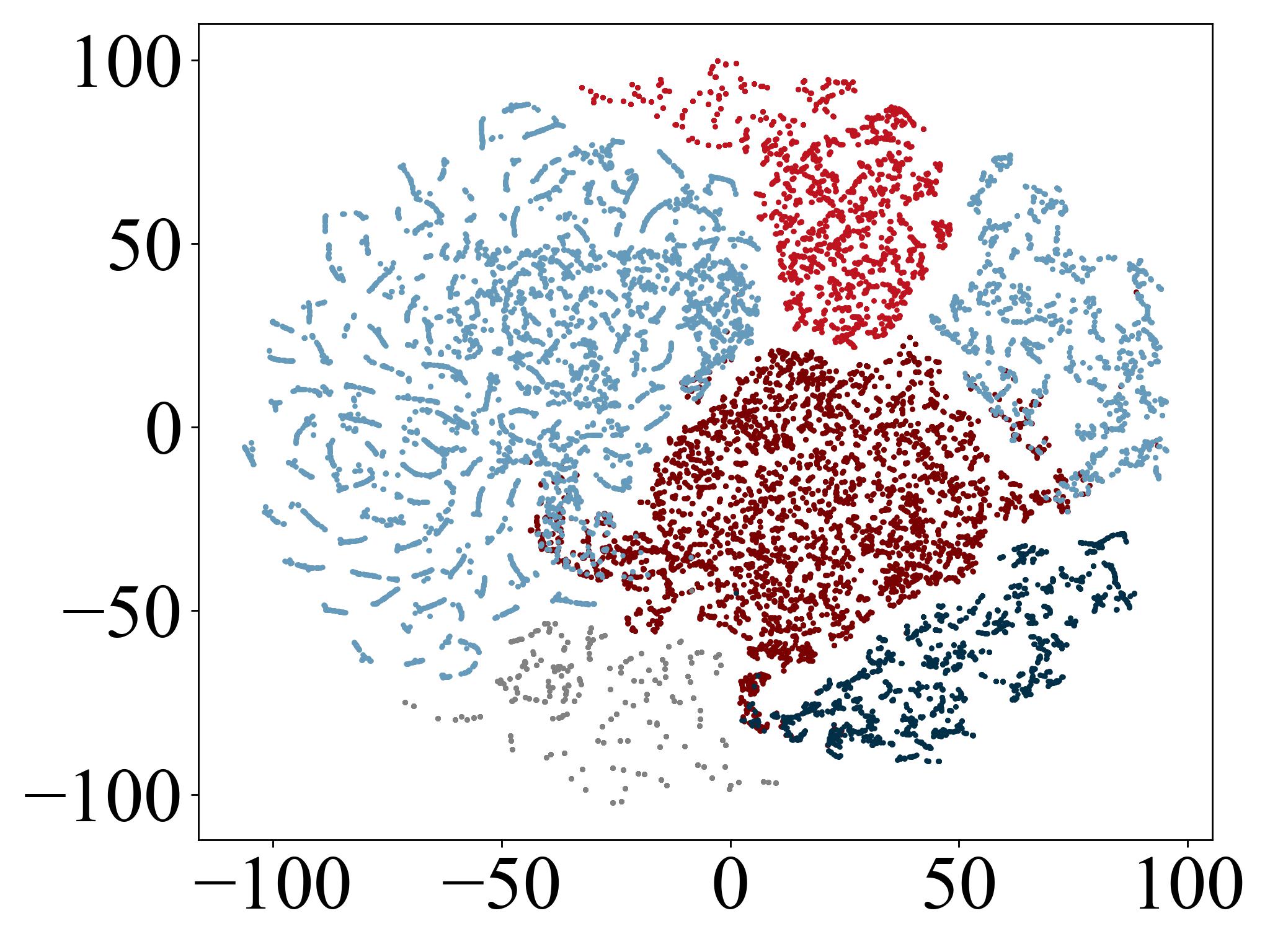}
\label{fig_5-1c}}
\subfloat[Epoch 150.]{\includegraphics[width=0.24\textwidth]{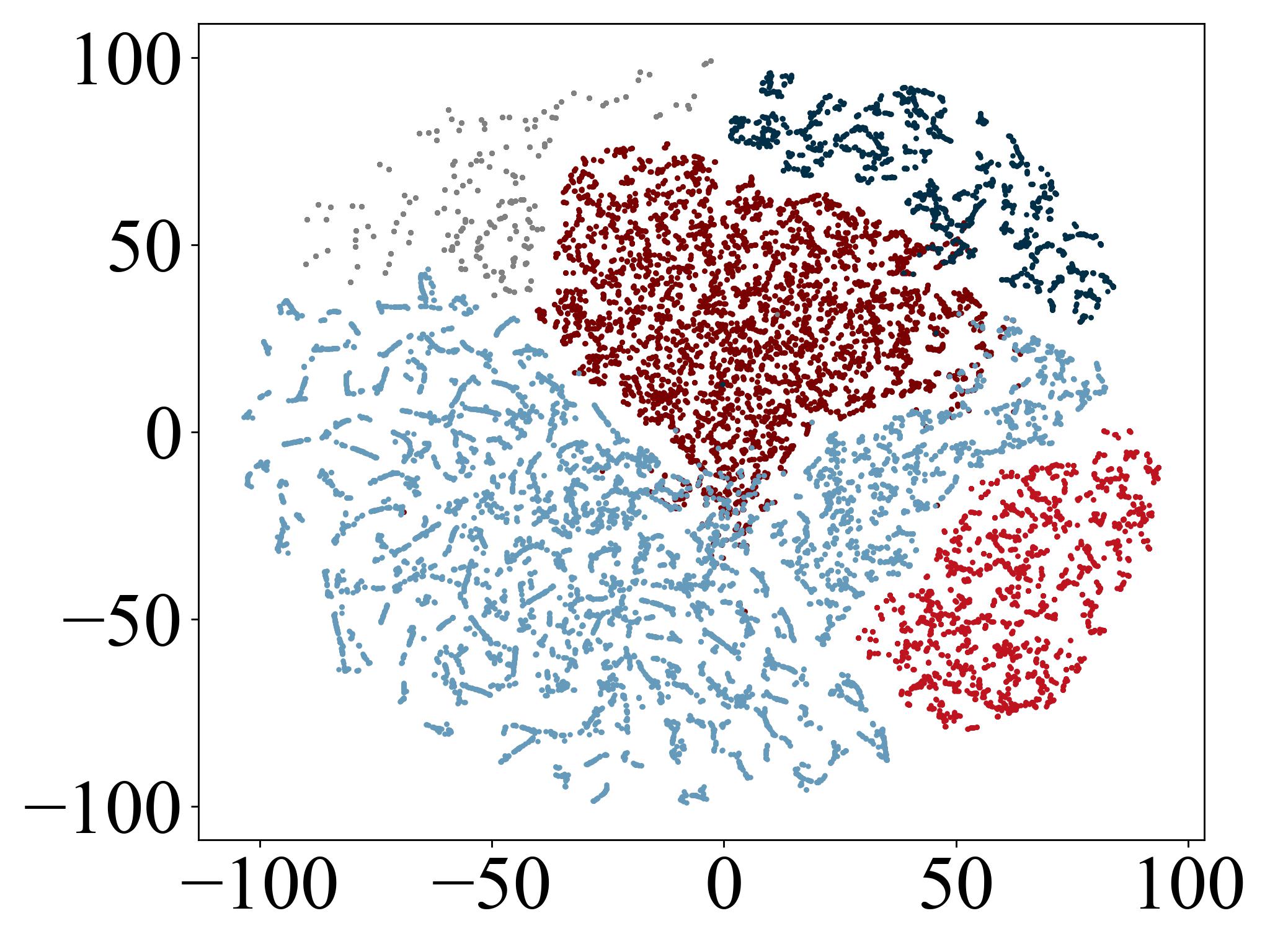}
\label{fig_5-1d}}
\caption{Visualization of embedded representations using t-SNE on the MAWI dataset, illustrating the evolution of the proposed method during training.}
\label{fig_5-1}
\end{figure*}

Fig. 4 (a) shows the t-SNE visualization of the original data space, where the results appear relatively disordered with no clear boundaries between classes. After one epoch of contrastive learning training, as shown in Fig. 4 (b), the data points begin to cluster gradually, with discernible separation between classes. By the 50th training epoch, as shown in Fig. 4 (c), intra-class features become more tightly clustered (e.g., the deep red region), while inter-class distances remain distinct. Following 150 epochs of training, as shown in Fig. 4 (d), the overall clustering is further enhanced, with classes exhibiting more compact distributions (e.g., light blue and light red regions) and clearer inter-class boundaries. This visualization demonstrates that the two components of the contrastive loss employed in the Embedding layer training—namely, enforcing intra-class distances to be less than the threshold $m$ and inter-class distances to exceed $2m$ — effectively fulfill their intended roles.

\subsection{Evaluation of Encrypted Traffic Classification}
With the increasing adoption and widespread use of encryption technologies, the invisibility of payload content has posed greater challenges to traffic classification tasks, making the evaluation of detection performance on encrypted traffic increasingly important. We extracted SSL-encrypted traffic from the data set collected on March 1, 2021, and carried out experiments on three different folds. The results are presented in Table IV.
\begin{table}[htbp]
    \centering
    \caption{EVALUATION OF ENCRYPTED TRAFFIC CLASSIFICATION ON THE MAWI DATASET}
    \begin{tabularx}{0.48\textwidth}{
        >{\centering\arraybackslash}p{0.09\textwidth} |
        >{\centering\arraybackslash}p{0.07\textwidth} |
        >{\centering\arraybackslash}X 
        >{\centering\arraybackslash}X 
        >{\centering\arraybackslash}X 
        }
        \toprule
        \textbf{Fold} & \textbf{Label} & \textit{Pre} & \textit{Rec} & \textit{F1} \\
        \midrule
        \multirow{2}{*}{1} & Benign & 92.23\% & 98.91\% & 95.46\% \\
        & Malicious & 98.35\% & 88.56\% & 93.20\% \\
        \cmidrule{1-5}
        \multirow{2}{*}{2} & Benign & 91.88\% & 98.19\% & 94.93\% \\
        & Malicious & 97.23\% & 88.00\% & 92.38\% \\
        \cmidrule{1-5}
        \multirow{2}{*}{3} & Benign & 90.88\% & 98.69\% & 94.63\% \\
        & Malicious & 97.97\% & 86.45\% & 91.85\% \\
        \bottomrule
    \end{tabularx}
    \label{tab:table5-3}
\end{table}

The experimental results show that \textit{FlowXpert} achieves recall rates of 98.91\%, 98.19\%, and 98.69\% for Benign traffic across the three folds, which are comparable to the recall rates observed in the general evaluation. This demonstrates the robustness and reliability of \textit{FlowXpert} in practical applications. On the other hand, the recall rates for malicious traffic across the three folds are 88.56\%, 88.00\% and 86.45\%, which, although slightly lower than the overall results, still represent a high level of performance considering the inherent challenges posed by encrypted traffic. We attribute \textit{FlowXpert}’s superior performance in detecting encrypted traffic to its feature extraction approach, which effectively integrates contextual information. This integration preserves the deep semantic features of network flows, thereby enhancing detection capabilities across multiple dimensions, even in complex encrypted scenarios.

\subsection{Evaluation of Model Generalization}
The generalization capability of traffic detection models is critical for real-world deployment, as the feature distribution of network traffic changes over time. Overfitted models may suffer significant performance drops in evolving environments, limiting their effectiveness. Given the high sensitivity and stringent reliability requirements of network security, ensuring accurate detection of benign traffic is essential. Even a 1\% performance drop can lead to substantial misclassification and disruption of normal services. To assess generalization, we trained models on two baseline dates, March 1, 2021 and June 3, 2023, and evaluated them at three subsequent weekly intervals. The 2021 model was tested on data from March 8, 15, and 22, while the 2023 model was tested on June 10, 18, and 25. The results are presented in Table V and Table VI.
\begin{table}[htbp]
    \centering
    \caption{GENERALIZATION EVALUATION ON THE MAWI DATASET (MARCH, 2021)}
    \begin{tabularx}{0.48\textwidth}{
        >{\centering\arraybackslash}p{0.09\textwidth} |
        >{\centering\arraybackslash}p{0.07\textwidth} |
        >{\centering\arraybackslash}X 
        >{\centering\arraybackslash}X 
        >{\centering\arraybackslash}X
        }
        \toprule
        \textbf{Date} & \textbf{Label} & \textit{Pre} & \textit{Rec} & \textit{F1} \\
        \midrule
        \multirow{2}{*}{\textit{2021/3/1}} & Benign & 98.98\% & 99.85\% & 99.41\% \\
        & Malicious & 99.12\% & 94.34\% & 96.67\% \\
        
        \midrule
        \multirow{2}{*}{\textit{2021/3/8}} & Benign & 89.58\% & 96.21\% & 92.78\% \\
        & Malicious & 69.59\% & 43.65\% & 53.65\% \\
        
        \midrule
        \multirow{2}{*}{\textit{2021/3/15}} & Benign & 86.41\% & 87.19\% & 86.80\% \\
        & Malicious & 48.80\% & 47.09\% & 47.93\% \\

        \midrule
        \multirow{2}{*}{\textit{\textbf{2021/3/22}}} & Benign & 92.59\% & 62.24\% & 74.44\% \\
        & Malicious & 10.81\% & 47.91\% & 17.64\% \\
        \bottomrule
    \end{tabularx}
    \label{tab:table5-4}
\end{table}

\begin{table}[htbp]
    \centering
    \caption{GENERALIZATION EVALUATION ON THE MAWI DATASET (JUNE, 2023)}
    \begin{tabularx}{0.48\textwidth}{
        >{\centering\arraybackslash}p{0.09\textwidth} |
        >{\centering\arraybackslash}p{0.07\textwidth} |
        >{\centering\arraybackslash}X 
        >{\centering\arraybackslash}X 
        >{\centering\arraybackslash}X
        }
        \toprule
        \textbf{Date} & \textbf{Label} & \textit{Pre} & \textit{Rec} & \textit{F1} \\
        \midrule
        \multirow{2}{*}{\textit{2023/6/3}} & Benign & 99.26\% & 99.32\% & 99.29\% \\
        & Malicious & 96.40\% & 96.07\% & 96.23\% \\
        
        \midrule
        \multirow{2}{*}{\textit{2023/6/10}} & Benign & 91.23\% & 89.00\% & 90.10\% \\
        & Malicious & 63.71\% & 69.30\% & 66.39\% \\
        
        \midrule
        \multirow{2}{*}{\textit{2023/6/18}} & Benign & 92.91\% & 92.83\% & 92.87\% \\
        & Malicious & 53.77\% & 54.07\% & 53.92\% \\

        \midrule
        \multirow{2}{*}{\textit{\textbf{2023/6/25}}} & Benign & 83.15\% & 84.41\% & 83.78\% \\
        & Malicious & 10.06\% & 9.25\% & 9.63\%  \\
        \bottomrule
    \end{tabularx}
    \label{tab:table5-5}
\end{table}

For the model trained on March 1, 2021, the recall rate for benign traffic on March 8 dropped slightly to 96.21\%, remaining within an acceptable range and ensuring operational stability, and the recall for malicious traffic fell to 43.65\%, indicating a noticeable decline but retaining some practical value. On March 15, 2021, benign recall decreased to 87. 19\%, and the malicious recall increased slightly to 47. 09\%, suggesting moderate robustness for malicious detection. On March 22, benign recall decreased significantly, indicating insufficient generalization for long-term deployment and the need for model retraining. This degradation over time reflects a common limitation of current AI models. Similarly, the model trained on June 3, 2023, achieved 89.00\% benign and 69.30\% malicious recall on June 10, maintaining strong utility. On June 18, 2023, benign recall improved to 92. 83\%, while malicious recall decreased to 54.00\%, showing some generalization capacity. However, by June 25, 2023, malicious recall plummeted to 9.00\%, signaling an urgent need for model updates.

In general, \textit{FlowXpert} demonstrates promising generalization to future data. Despite moderate drops in malicious recall within two weeks, its performance is notable given the limited training data (~2 hours). Increasing the training sample size is expected to further enhance the temporal robustness. Importantly, \textit{FlowXpert} consistently maintains high benign recall, minimizing false alarms, and ensuring sustained usability in real-world deployment.

\subsection{Comparison with the SOTA Methods}
\begin{figure*}[htbp]
\centering
\subfloat[Precision.]{\includegraphics[width=0.28\textwidth]{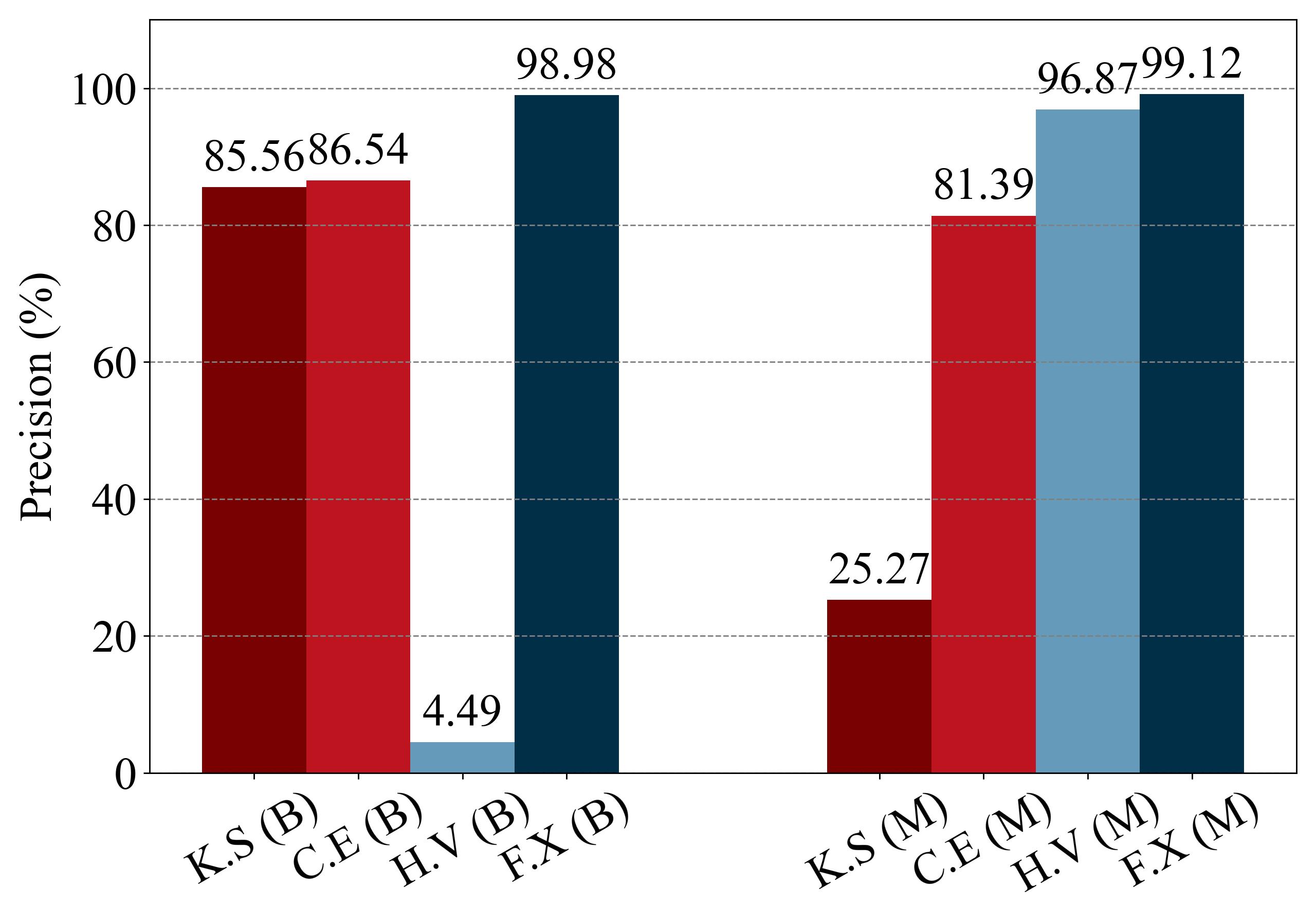}
\label{fig_5-2a}}
\hspace{0.03\textwidth}
\subfloat[Recall.]{\includegraphics[width=0.28\textwidth]{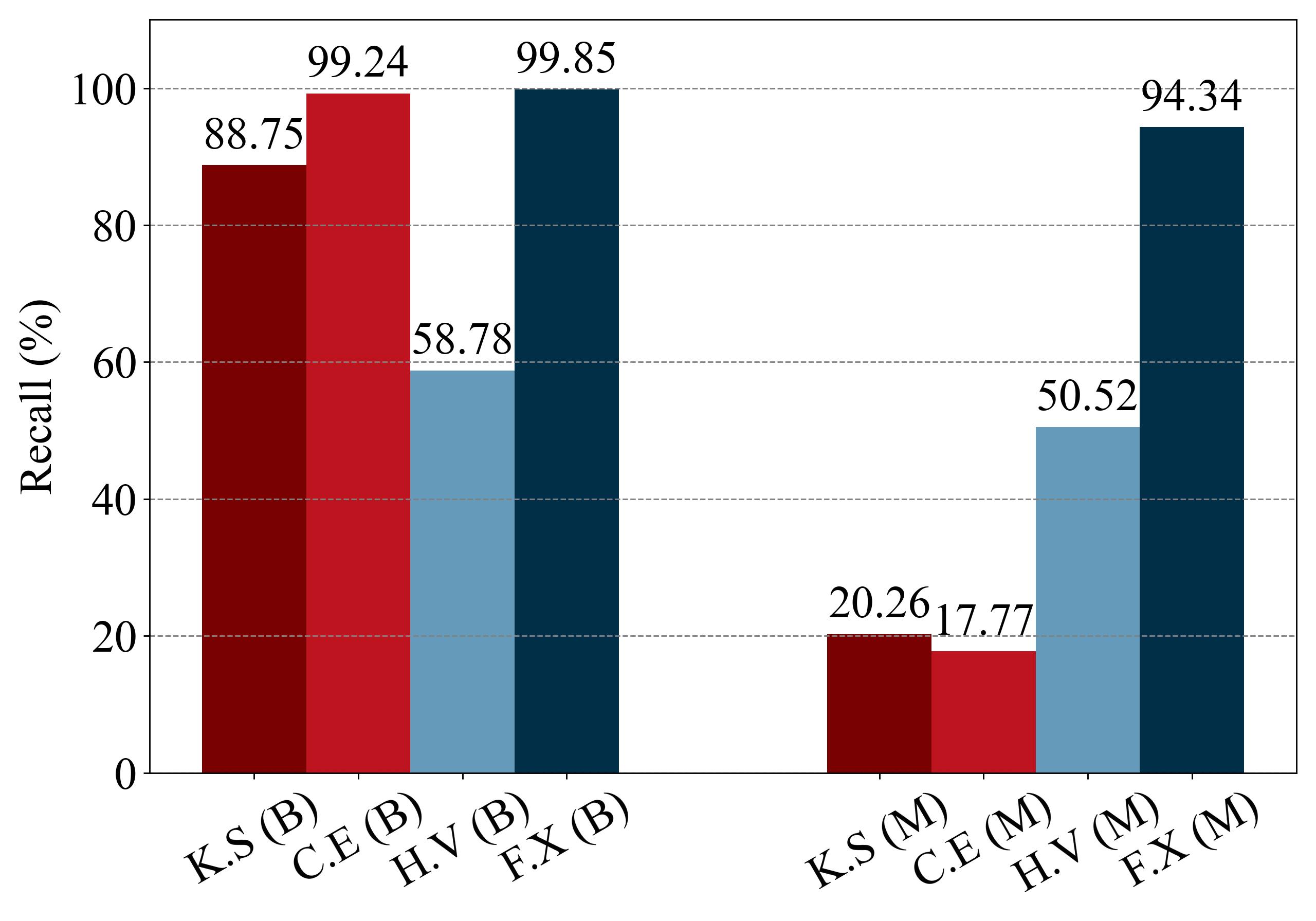}
\label{fig_5-2b}}
\hspace{0.03\textwidth}
\subfloat[F1 score.]{\includegraphics[width=0.28\textwidth]{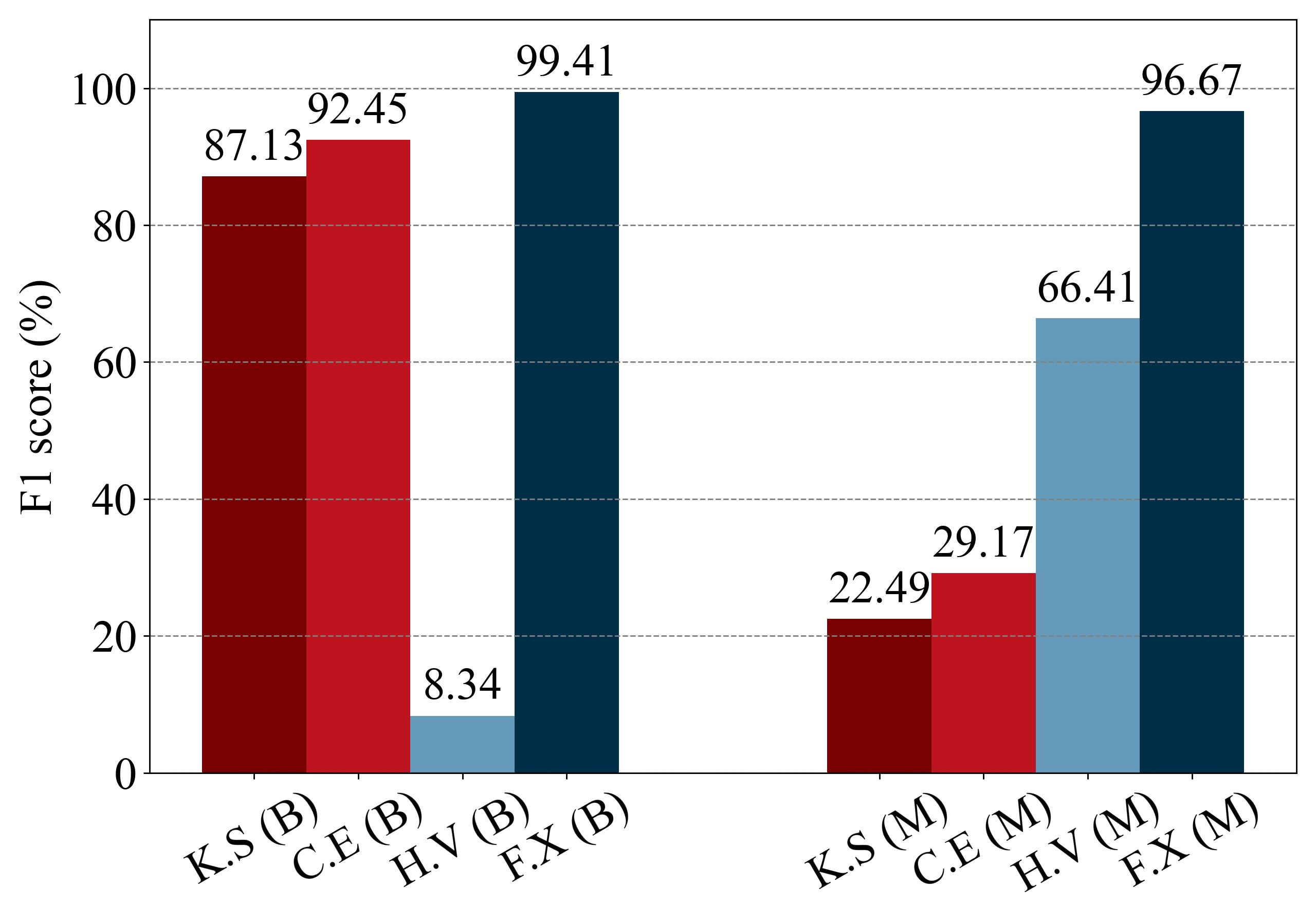}
\label{fig_5-2c}}
\caption{Comparative experiments with SOTA methods on the MAWI dataset (2021/3/1).}
\label{fig_5-2}
\end{figure*}

\begin{figure*}[htbp]
\centering
\subfloat[Precision.]{\includegraphics[width=0.28\textwidth]{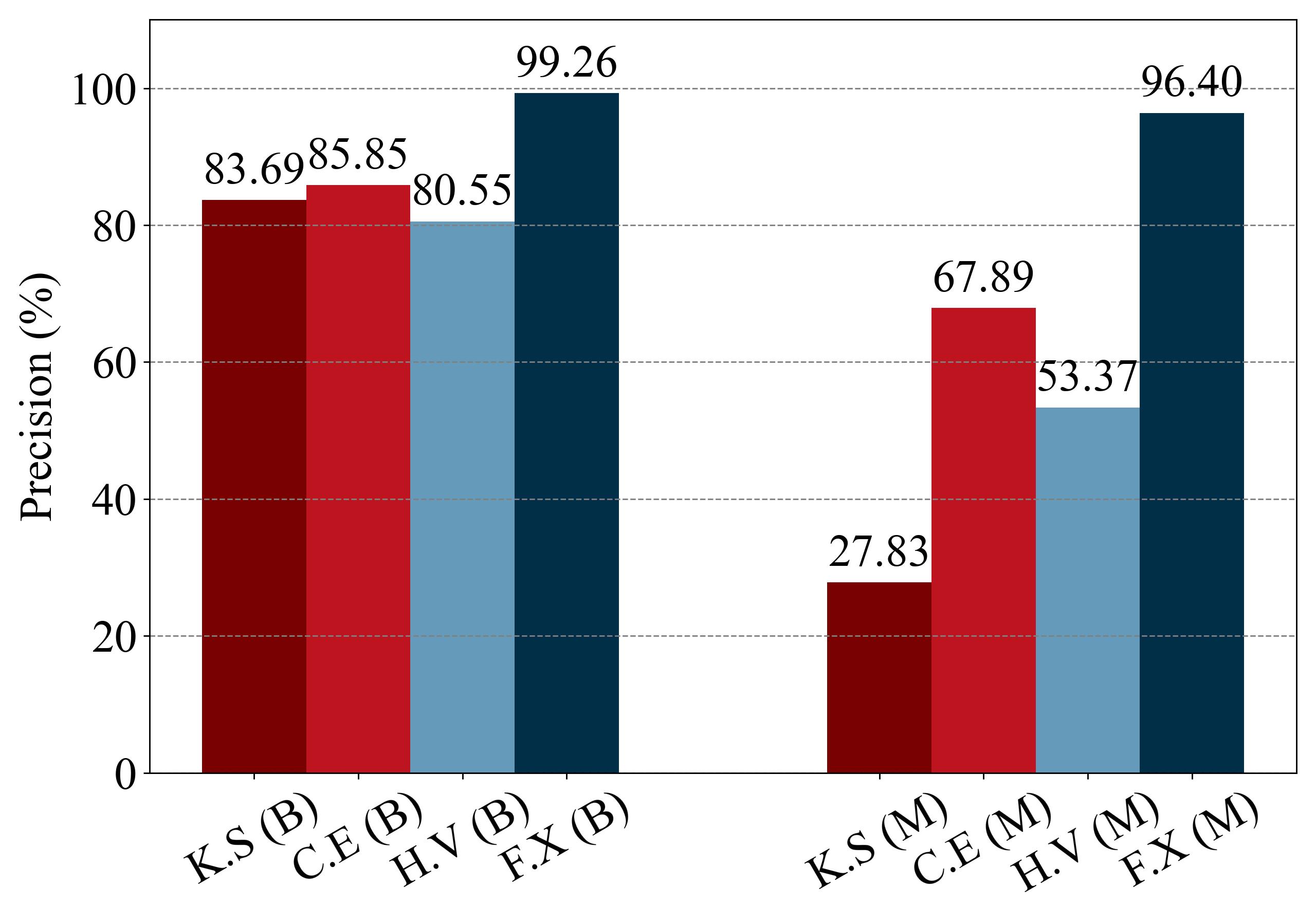}
\label{fig_5-3a}}
\hspace{0.03\textwidth}
\subfloat[Recall.]{\includegraphics[width=0.28\textwidth]{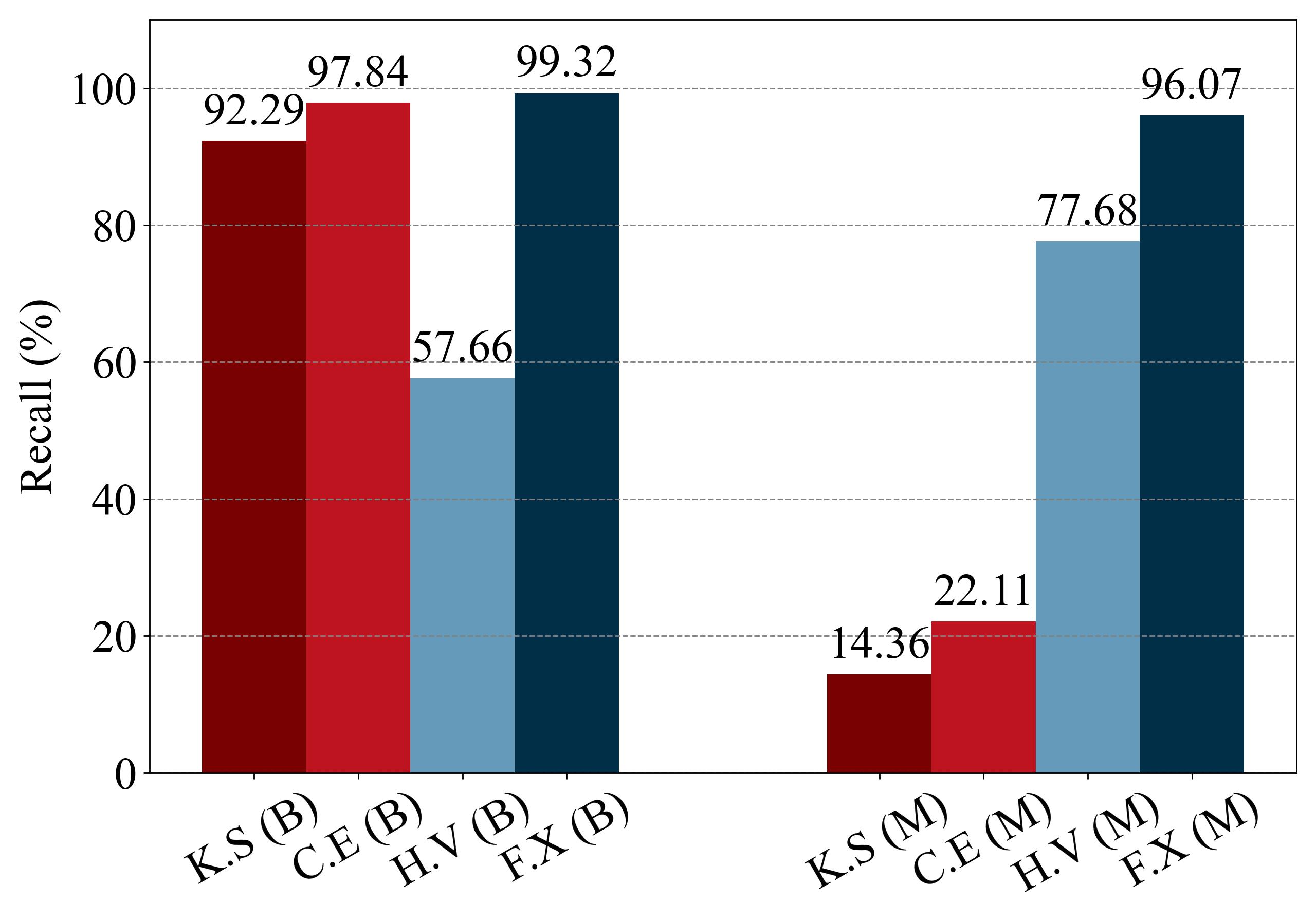}
\label{fig_5-3b}}
\hspace{0.03\textwidth}
\subfloat[F1 score.]{\includegraphics[width=0.28\textwidth]{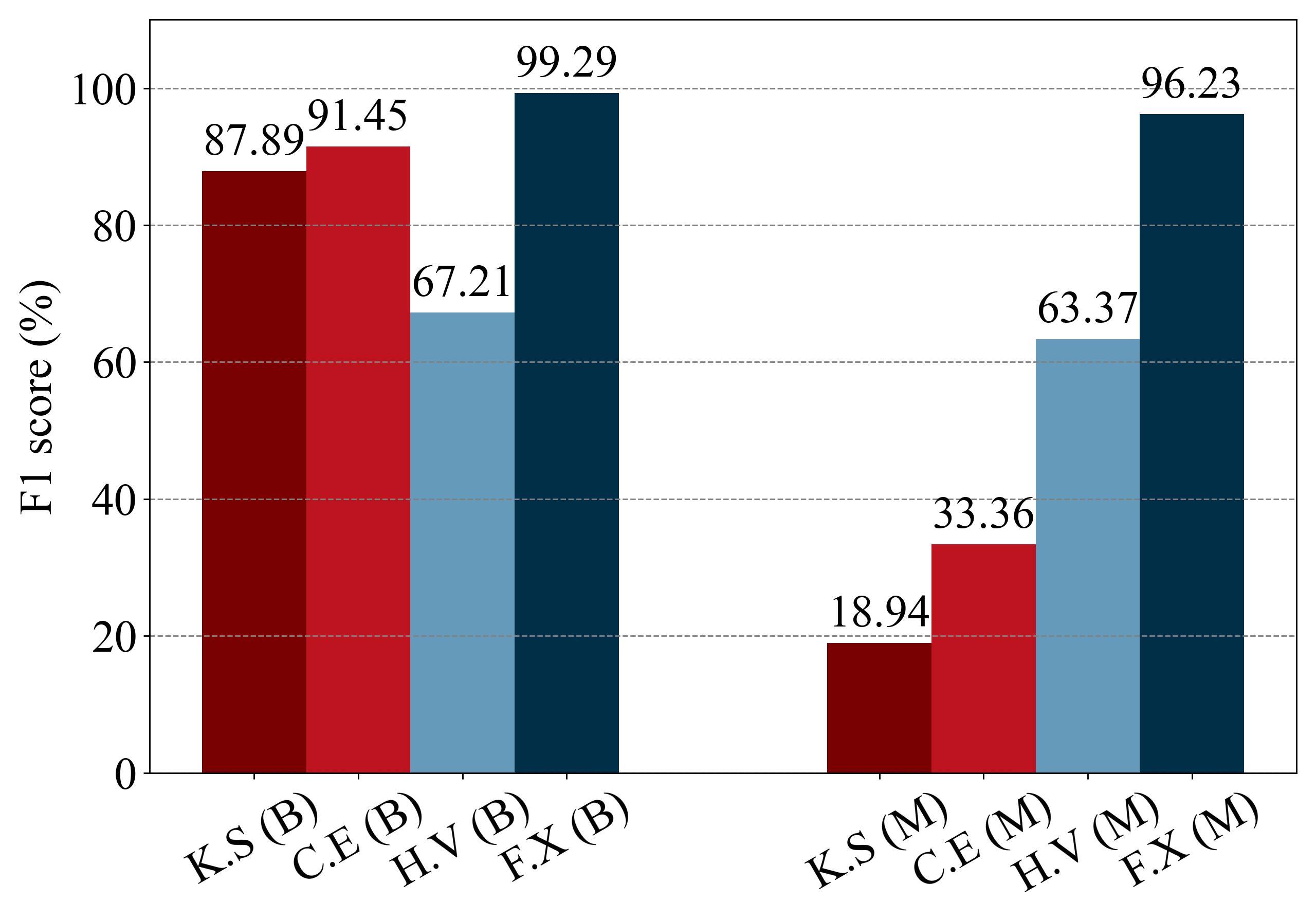}
\label{fig_5-3c}}
\caption{Comparative experiments with SOTA methods on the MAWI dataset (2023/6/3).}
\label{fig_5-3}
\end{figure*}
To highlight the contribution of \textit{FlowXpert} in traffic detection, we compared it with several recent SOTA methods. Specifically, we selected \textit{Kitsune} (K.S) \cite{mirsky2018kitsune}, a classic and widely discussed approach; \textit{CVAE-EVT} (C.E) \cite{yang2021conditional}, a probabilistic classification method; and \textit{HyperVision} (H.V) \cite{fu2023detecting}, designed for encrypted traffic detection. These methods were chosen for their representative status, rigorous theoretical underpinnings, and comprehensive analyses. The comparison results are presented in Fig. 5 and Fig. 6, where B and M in parentheses denote Benign and Malicious, respectively.

Firstly, as illustrated in Fig. 5 (b), in the data set corresponding to March 1, 2021, \textit{Kitsune}, \textit{CVAE-EVT}, and our proposed \textit{FlowXpert} all demonstrate strong performance in terms of recall for normal traffic detection, with \textit{CVAE-EVT} and \textit{FlowXpert} both achieving over 99\%, whereas \textit{HyperVision} shows a relatively poor recall of 58.78\%. However, for malicious traffic detection, performance disparities become more pronounced. \textit{FlowXpert} achieves a recall of 94.34\%, significantly outperforming all baseline methods. The best-performing baseline, \textit{HyperVision}, reaches only 50.52\%, while \textit{Kitsune} and \textit{CVAE-EVT} both hover around 20\%, indicating a substantial gap. Furthermore, as shown in Fig. 5 (a) and Fig. 5 (c), \textit{FlowXpert} also leads in terms of precision and F1 score, consistently outperforming the other compared methods.

Secondly, as shown in Fig. 6 (b), in the data set corresponding to June 3, 2023, \textit{FlowXpert} achieves a recall of 99.32\% for normal traffic, closely followed by \textit{CVAE-EVT} (97.84\%) and \textit{Kitsune} (92.29\%). The difference is relatively small among these three methods. \textit{HyperVision} lags significantly, with a recall of only 57.66\% for normal traffic. For malicious traffic detection, \textit{FlowXpert} again demonstrates a clear advantage, achieving a recall of 96.07\%, substantially outperforming \textit{HyperVision} (77.68\%), \textit{CVAE-EVT} (22.11\%), and \textit{Kitsune} (14.36\%). This highlights \textit{FlowXpert}’s superior ability to detect anomalous behavior in challenging real-world settings. Furthermore, as shown in Fig. 6 (a) and Fig. 6 (c), \textit{FlowXpert} also outperforms all baselines in terms of precision and F1 score, further confirming its overall effectiveness and robustness in both benign and malicious traffic scenarios.

In summary, \textit{Kitsune} and \textit{CVAE-EVT} exhibit strong performance in detecting benign traffic, with results comparable to those of \textit{FlowXpert}. However, their malicious traffic detection performance is significantly inferior. This can be attributed to their reliance on traditional flow-level features as model input during training and inference. As discussed in Section III, the inherent sparsity of flow features negatively affects model convergence, ultimately leading to degraded performance in real-world scenarios. Although \textit{HyperVision} incorporates graph-based interaction features, it constructs interaction graphs using IP addresses as nodes. This design becomes problematic in practice, where the number of unique IPs is often large. In such cases, clustering algorithms may not be suitable for detection tasks, and more importantly, the memory required to maintain the interaction graph grows rapidly, posing significant challenges for practical deployment. \textit{FlowXpert} addresses both of these critical issues. Firstly, it mitigates the convergence challenges of flow-based features by selecting only a small subset of flow features and incorporating contextual features related to the source host, resulting in reduced memory consumption and improved detection capability. Second, by integrating contrastive learning with the DBSCAN unsupervised clustering algorithm, \textit{FlowXpert} constructs robust traffic embeddings, further enhancing detection performance.

\subsection{Evaluation of Real-Time Performance}
Given the stringent real-time requirements and limited device resources in IoT scenarios, we conducted latency and throughput tests on a low-end machine to evaluate the practicality of our method. The hardware configuration is summarized as follows:  a 13th Gen Intel® Core™ i5-13500 CPU, Windows 11, 32GB DDR5 5200Hz RAM, and 512GB SSD. The results are presented in Fig. 7.
\begin{figure}[htbp]
\centering
\subfloat[Embedding.]{\includegraphics[width=0.24\textwidth]{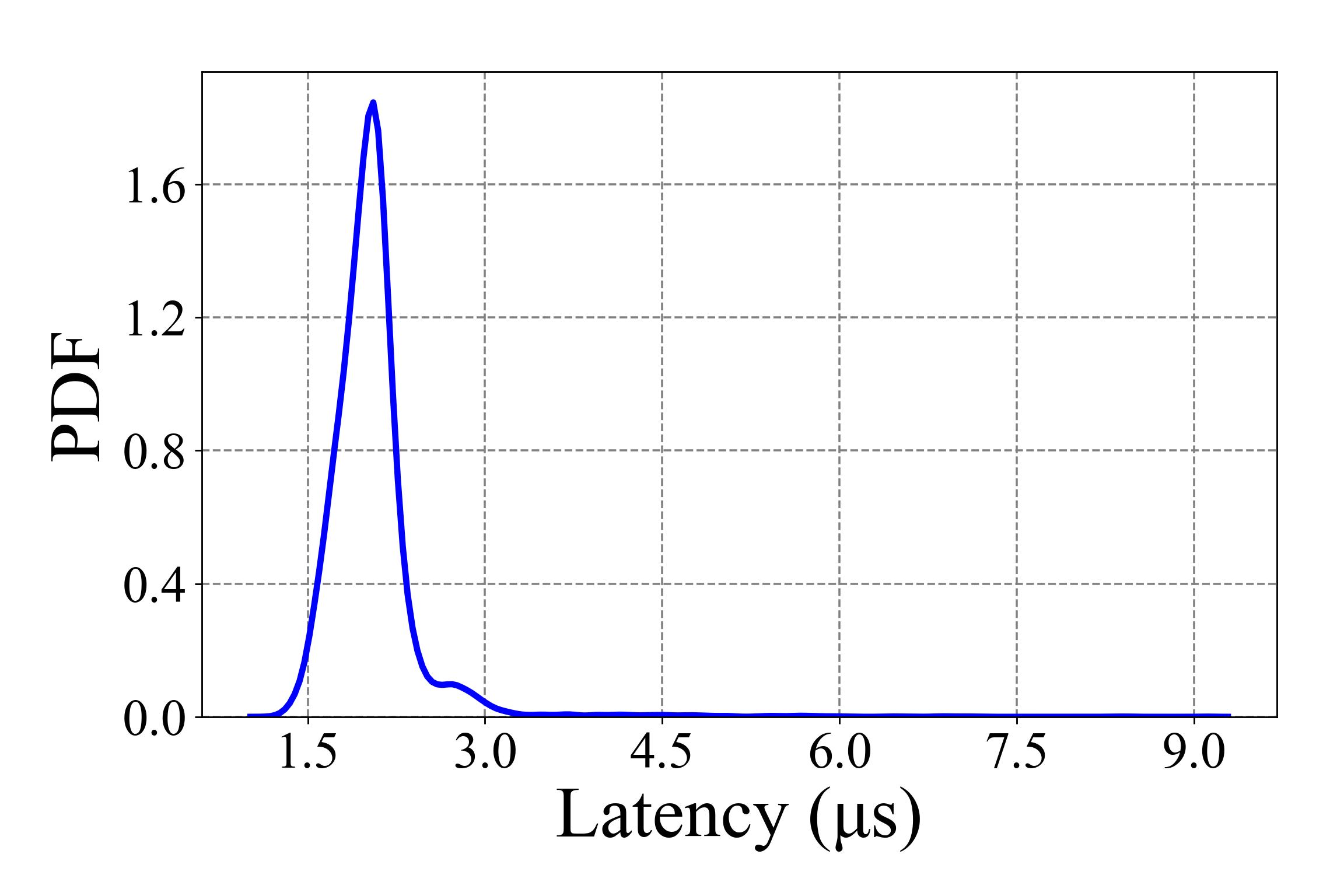}
\label{fig_5-4a}}
\subfloat[Embedding.]{\includegraphics[width=0.24\textwidth]{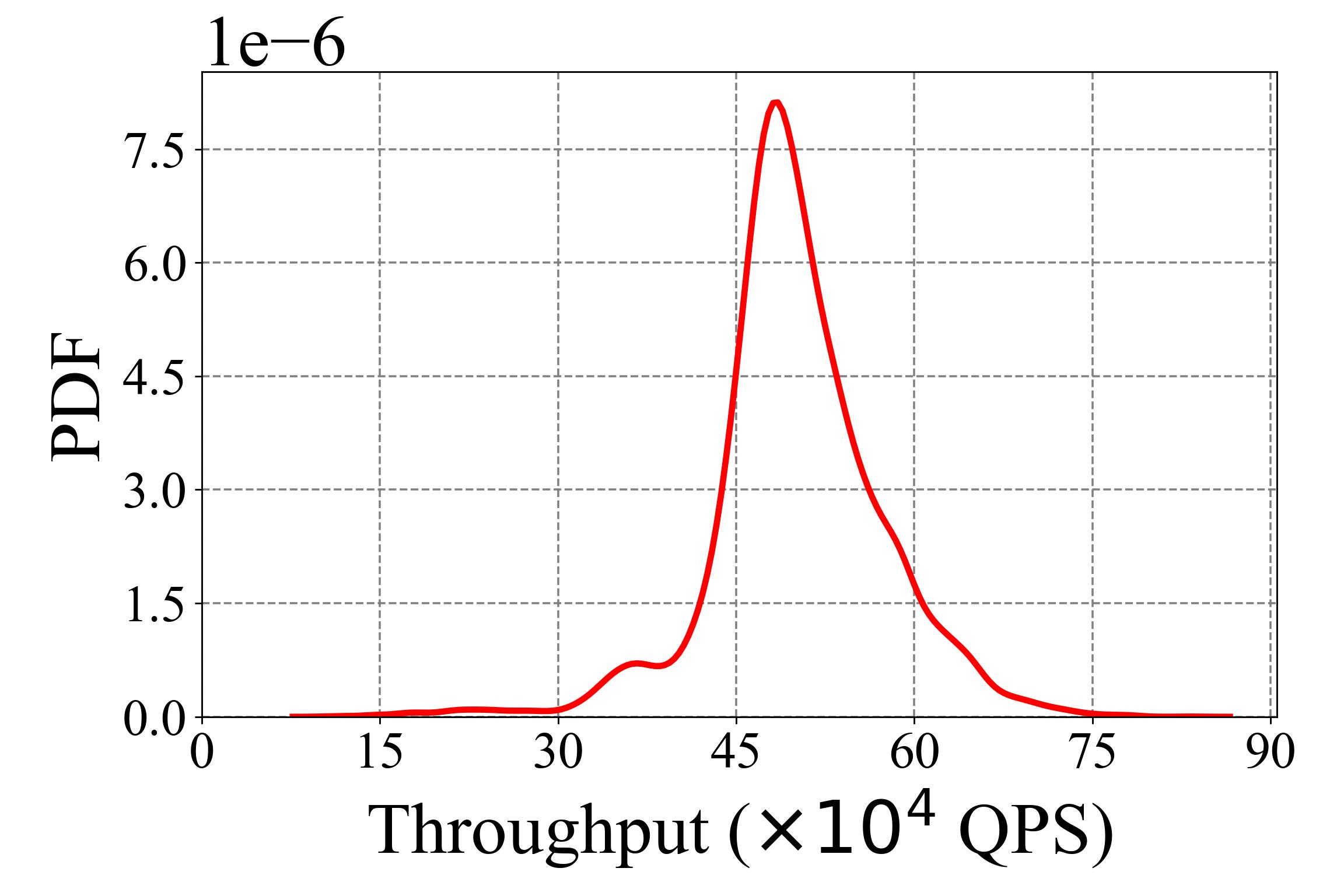}
\label{fig_5-4b}}

\subfloat[Encoder.]{\includegraphics[width=0.24\textwidth]{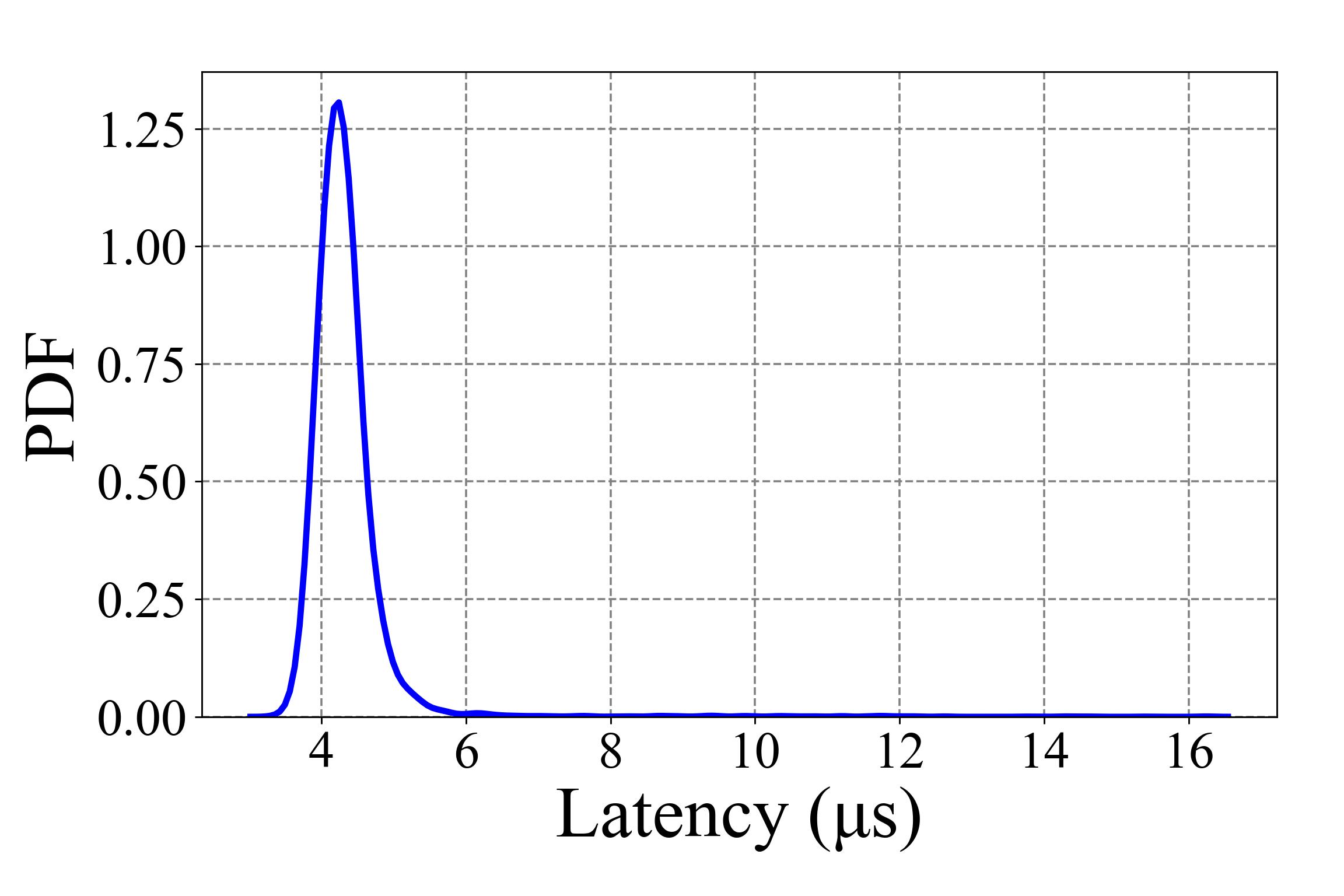}
\label{fig_5-4c}}
\subfloat[Encoder.]{\includegraphics[width=0.24\textwidth]{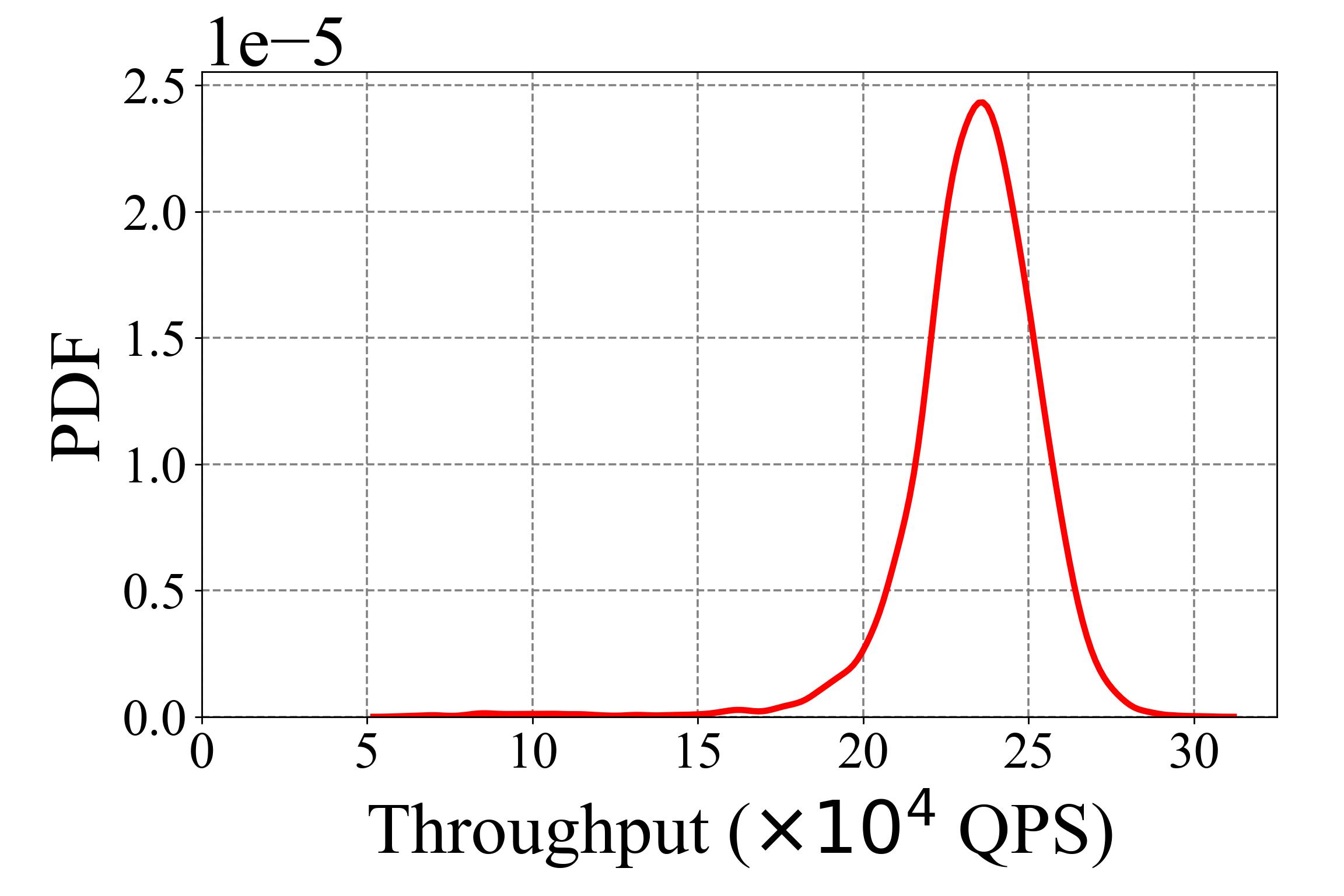}
\label{fig_5-4d}}

\subfloat[FlowXpert.]{\includegraphics[width=0.24\textwidth]{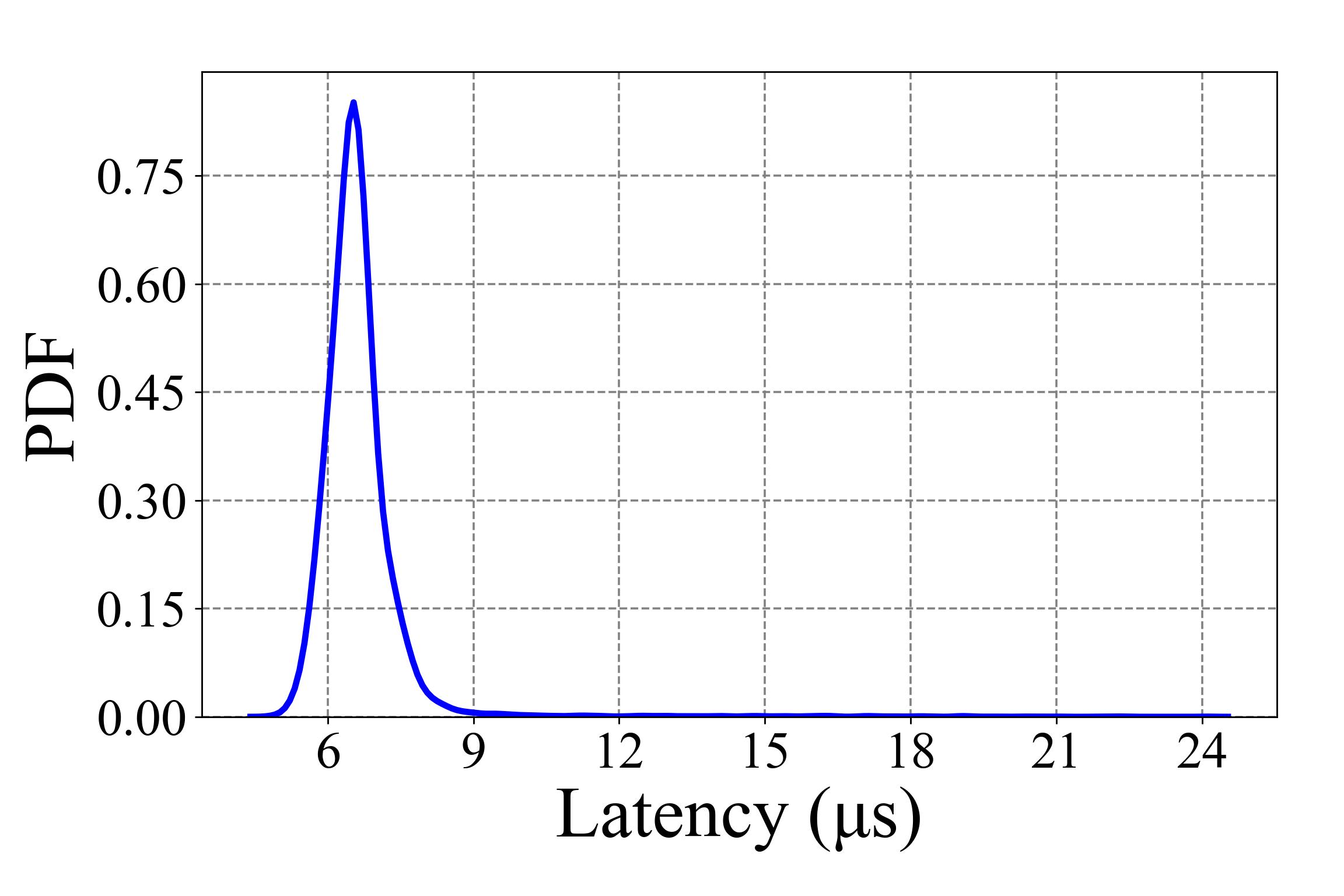}
\label{fig_5-4e}}
\subfloat[FlowXpert.]{\includegraphics[width=0.24\textwidth]{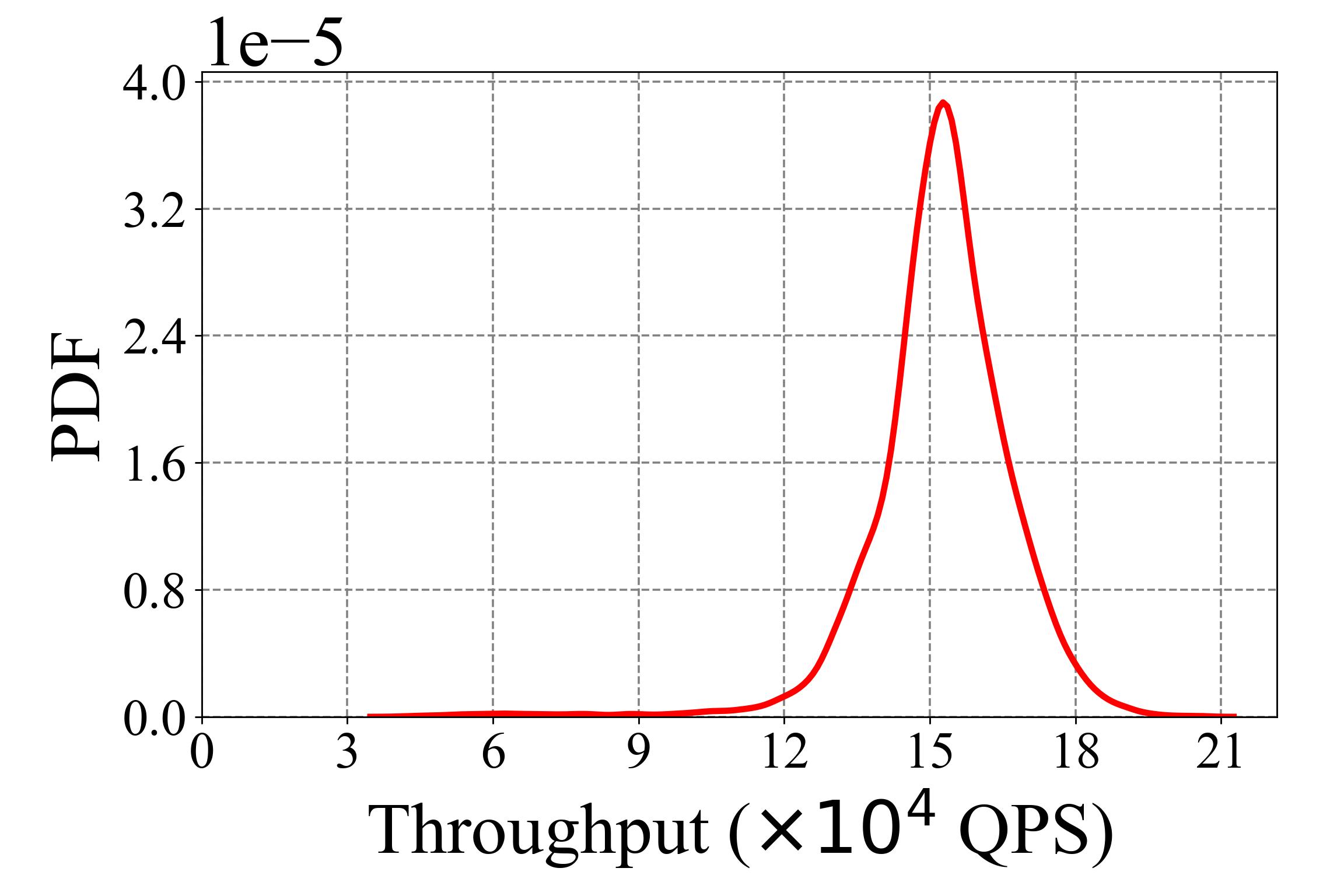}
\label{fig_5-4f}}
\caption{Performance test of FlowXpert on the MAWI dataset.}
\label{fig_5-4}
\end{figure}

As shown in Fig. 7 (a) and Fig. 7 (b), \textit{FlowXpert} achieves an average latency of approximately 2 microseconds on the embedding component, with an average throughput of 500,000 queries per second (QPS). In Fig. 7 (c) and Fig. 7 (d), the encoder exhibits an average latency of around 4 microseconds and an average throughput of 250,000 QPS. Fig. 7 (e) and Fig. 7 (f) present the overall performance, where the average end-to-end latency remains below 7 microseconds, and the throughput averages around 160,000 QPS. In general, these results demonstrate that \textit{FlowXpert} is well-suited for deployment in real-time environments.

Furthermore, to quantify the computational footprint of the proposed model, we analyze the number of parameters and corresponding memory consumption (see Eq. (19), Eq. (20) and Eq. (21)). The Embedding module comprises approximately 4.4K parameters, while the Encoder module contains around 180K parameters. The final Classifier adds another 258 parameters. In total, the model contains 185,234 parameters. Assuming that each parameter is stored using 32-bit floating point precision (float32), the total memory consumption amounts to approximately 724 KB. This lightweight design ensures that the model is suitable for deployment in resource-constrained environments such as edge devices or IoT devices.

\section{Related Work}
In this section, we provide a discussion and summary of recent significant studies on ML-based algorithms and graph-based methods for traffic detection tasks. These comparisons are further illustrated in Table VIII.

\begin{table*}[htbp]
  \centering
  \caption{COMPARISON OF RECENT INTRUSION DETECTION TECHNIQUES.}
    \begin{tabularx}{0.96\textwidth}{
        >{\centering\arraybackslash}p{0.12\textwidth} | 
        >{\centering\arraybackslash}p{0.24\textwidth} |
        >{\centering\arraybackslash}X 
        >{\centering\arraybackslash}X 
        >{\centering\arraybackslash}X |
        >{\centering\arraybackslash}X
        >{\centering\arraybackslash}X
    }
    \toprule
     \multirow{3}{*}{Methods} & \multirow{3}{*}{Used Technology} & \multicolumn{3}{c|}{Design Goals} & \multicolumn{2}{c}{Detection Performance} \\ 
     \cmidrule{3-7}
     & & Encrypted Traffic & Realtime Detection & \multirow{2}{*}{Generalization} & \multirow{2}{*}{Low Latency} & High Throughput \\
    \midrule
    Kitsune \cite{mirsky2018kitsune} & ensemble of autoencoders & \textcolor{red}{\texttimes} & \textcolor{blue}{\checkmark} & \textcolor{red}{\texttimes} & \textcolor{red}{\texttimes} & \textcolor{red}{\texttimes} \\
    SCAE-SVM \cite{wang2020cloud} & autoencoder, support vector machine& \textcolor{red}{\texttimes} & \textcolor{red}{\texttimes} & \textcolor{red}{\texttimes} & \textcolor{red}{\texttimes} & \textcolor{red}{\texttimes} \\
    CVAE-EVT \cite{yang2021conditional} & autoencoder, extreme value theory & \textcolor{red}{\texttimes} & \textcolor{red}{\texttimes} & \textcolor{red}{\texttimes} & \textcolor{red}{\texttimes} & \textcolor{red}{\texttimes} \\
    Whisper \cite{fu2021realtime} & frequency feature, clustering & \textcolor{red}{\texttimes} & \textcolor{blue}{\checkmark} & \textcolor{red}{\texttimes} & \textcolor{blue}{\checkmark} & \textcolor{blue}{\checkmark} \\
    DM-IDS \cite{zha2025dm} & dual-modal fusion & \textcolor{red}{\texttimes} & \textcolor{blue}{\checkmark} & \textcolor{red}{\texttimes} & \textcolor{blue}{\checkmark} & \textcolor{blue}{\checkmark} \\
    DeepLog \cite{du2017deeplog} & long short-term memory, log & \textcolor{red}{\texttimes} & \textcolor{red}{\texttimes} & \textcolor{red}{\texttimes} & \textcolor{red}{\texttimes} & \textcolor{red}{\texttimes} \\
    DynaMiner \cite{eshete2017dynaminer} & web conversation graph analytics & \textcolor{red}{\texttimes} & \textcolor{blue}{\checkmark} & \textcolor{red}{\texttimes} & \textcolor{red}{\texttimes} & \textcolor{red}{\texttimes} \\
    Nazca \cite{invernizzi2014nazca} & neighborhood graph & \textcolor{red}{\texttimes} & \textcolor{blue}{\checkmark} & \textcolor{red}{\texttimes} & \textcolor{red}{\texttimes} & \textcolor{red}{\texttimes} \\
    HyperVision \cite{fu2023detecting} & interaction graph, clustering & \textcolor{blue}{\checkmark} & \textcolor{blue}{\checkmark} & \textcolor{red}{\texttimes} & \textcolor{blue}{\checkmark} & \textcolor{blue}{\checkmark} \\
    EdgeTorrent \cite{king2023edgetorrent} & provenance graph, transformer & \textcolor{red}{\texttimes} & \textcolor{blue}{\checkmark} & \textcolor{red}{\texttimes} & \textcolor{blue}{\checkmark} & \textcolor{blue}{\checkmark} \\
    TCG-IDS \cite{wu2025tcg} & graph learning, contrastive learning & \textcolor{red}{\texttimes} & \textcolor{blue}{\checkmark} & \textcolor{red}{\texttimes} & \textcolor{red}{\texttimes} & \textcolor{red}{\texttimes} \\
    
    \textbf{\textit{FlowXpert}} & context-aware, contrastive learning & \textcolor{blue}{\checkmark} & \textcolor{blue}{\checkmark} & \textcolor{blue}{\checkmark} & \textcolor{blue}{\checkmark} & \textcolor{blue}{\checkmark}\\
    \bottomrule
    \end{tabularx}%
  \label{tab:table_8-1}%
\end{table*}%

\paragraph{\textbf{ML Based Traffic Detection}} Traditional ML algorithms have been widely applied to network traffic detection tasks. For example, Mirsky et al. \cite{mirsky2018kitsune} introduced a unsupervised online network intrusion detection method, whose core component, KitNET, employs an ensemble architecture of small autoencoders to perform reconstruction error analysis on extracted network traffic features, thereby allowing the detection of anomalous behaviors, and Wang et al. \cite{wang2020cloud} proposed a cloud intrusion detection method based on a stacked contractive autoencoder combined with a Support Vector Machine. Yang et al. \cite{yang2021conditional} proposed a two-stage learning approach that integrates conditional variational autoencoders with extreme value theory to construct a hierarchical intrusion detection system. Moreover, by incorporating the clustering of benign traffic, the approach significantly reduces the false-positive rate. Fu et al. \cite{fu2021realtime} presented a method that uses features in the frequency domain to capture the temporal characteristics of attack traffic without losing contextual information and integrates clustering algorithms to achieve intrusion detection. Furthermore, Zha et al. \cite{zha2025dm} proposed an intrusion detection method that combines flow-based features and payload-based features through bilinear fusion, in order to enhance the accuracy and robustness of network intrusion detection, and Du et al. \cite{du2017deeplog} proposed using long short-term memory neural networks to treat system logs as natural language sequences, learning the patterns of log sequences under normal execution to detect anomalies effectively.

Traditional ML-based approaches are mostly trained in flow-level features, and while they often achieve impressive results on synthetic benchmark datasets, this focus on benchmark performance may overlook real-world practicalities \cite{sommer2010outside}. In realistic network environments, the sparsity of flow features can significantly hinder model convergence, ultimately rendering such methods ineffective. Moreover, conventional ML methods typically do not incorporate contextual information, which can lead to poor generalization and limit their applicability. In contrast, our proposed \textit{FlowXpert} leverages novel context-aware features as model input, deliberately discarding most traditional flow features. In addition, it integrates an unsupervised embedding training strategy to further enhance detection performance, offering a more robust and practical solution for real-world deployment.

\paragraph{\textbf{Graph Based Traffic Detection}} In recent years, graph-based algorithms have been increasingly applied to traffic detection. For example, Eshete et al. \cite{eshete2017dynaminer} constructed HTTP interaction graphs to detect malicious static resources, and Invernizzi et al. \cite{invernizzi2014nazca} used graphs derived from plaintext traffic to identify malicious infrastructure associated with malware. Fu et al. \cite{fu2023detecting} introduced the construction of an in-memory flow interaction graph combined with unsupervised graph learning to detect encrypted malicious traffic. King et al. \cite{king2023edgetorrent} proposed EdgeTorrent, which enables embedding of streaming in real time and anomaly detection for temporal provenance graphs. Wu et al. \cite{wu2025tcg} proposed a self-supervised temporal graph neural network for intrusion detection. The model constructs sequential temporal attributed graphs and applies temporal, asymmetric, and masked contrastive strategies to learn robust node representations, thereby enabling the detection of unknown attacks.

Traffic detection algorithms based on graph features can improve robustness, generalization, and overall performance. However, constructing such graphs is often complex and may hinder the practical deployment of these approaches. For example, when IP addresses are used as graph nodes, the memory graph can grow rapidly with increasing number of IPs, leading to a sharp increase in the number of subconnected components \cite{fu2023detecting}. Given the large number of IPs in real-world scenarios, this challenge is almost unavoidable and significantly limits the scalability of graph-based methods. In contrast, \textit{FlowXpert} adopts a more practical approach by considering only the contextual information of the source host and using unsupervised learning to train the model. This design not only improves the generalizability of the model, but also improves its practicality for real-world deployment.

\section{Conclusion}
We propose a novel traffic detection method for IoT environments, named \textit{FlowXpert}. First, we identify theoretical limitations in the widely adopted flow-based features used in existing detection methods, which can severely hinder model convergence. To address this, we introduce a new context-aware feature extraction method, termed FlowVision. Second, to improve generalization capability, we design an unsupervised embedding approach that integrates DBSCAN and contrastive learning for effective representation learning. Finally, we conduct extensive experiments on the real-world MAWI dataset to evaluate the detection performance, generalizability, and improvements over SOTA methods. Considering the real-time requirements of IoT scenarios, our model is lightweight and achieves ultralow detection latency, with per-sample inference time as low as 7 microseconds on a low-resource device.

\section*{Acknowledgments}
This work is supported by the Key Research and Development Program of Zhejiang Province (No. 2024SSYS0001).

\bibliographystyle{IEEEtran}
\bibliography{reference}

\vfill

\end{document}